\newif\iflandscape
\newif\ifportrait
\portraittrue
\newlength{\extralineskip}
\ifportrait
  \documentstyle[12pt]{article}
\fi
\iflandscape
  \documentstyle[twocolumn]{article}
\fi
\def\tr#1{{\rm tr}\kern-3pt\left[#1\right]}

\makeatletter
\newdimen\normalarrayskip 
\newdimen\minarrayskip 
\normalarrayskip\baselineskip
\minarrayskip\jot
\newif\ifold \oldtrue \def\new{\oldfalse}
\def\arraymode{\ifold\relax\else\displaystyle\fi} 
\def\eqnumphantom{\phantom{(\theequation)}} 
\def\@arrayskip{\ifold\baselineskip\z@\lineskip\z@
     \else
     \baselineskip\minarrayskip\lineskip2\minarrayskip\fi}
\def\@arrayclassz{\ifcase \@lastchclass \@acolampacol \or
\@ampacol \or \or \or \@addamp \or
   \@acolampacol \or \@firstampfalse \@acol \fi
\edef\@preamble{\@preamble
  \ifcase \@chnum
     \hfil$\relax\arraymode\@sharp$\hfil
     \or $\relax\arraymode\@sharp$\hfil
     \or \hfil$\relax\arraymode\@sharp$\fi}}
\def\@array[#1]#2{\setbox\@arstrutbox=\hbox{\vrule
     height\arraystretch \ht\strutbox
     depth\arraystretch \dp\strutbox
     width\z@}\@mkpream{#2}\edef\@preamble{\halign \noexpand\@halignto
\bgroup \tabskip\z@ \@arstrut \@preamble \tabskip\z@ \cr}%
\let\@startpbox\@@startpbox \let\@endpbox\@@endpbox
  \if #1t\vtop \else \if#1b\vbox \else \vcenter \fi\fi
  \bgroup \let\par\relax
  \let\@sharp##\let\protect\relax
  \@arrayskip\@preamble}
\def\eqnarray{\stepcounter{equation}%
              \let\@currentlabel=\theequation
              \global\@eqnswtrue
              \global\@eqcnt\z@
              \tabskip\@centering
              \let\\=\@eqncr
              $$%
 \halign to \displaywidth\bgroup
    \eqnumphantom\@eqnsel\hskip\@centering
    $\displaystyle \tabskip\z@ {##}$%
    &\global\@eqcnt\@ne \hskip 2\arraycolsep
         $\displaystyle\arraymode{##}$\hfil
    &\global\@eqcnt\tw@ \hskip 2\arraycolsep
         $\displaystyle\tabskip\z@{##}$\hfil
         \tabskip\@centering
    &{##}\tabskip\z@\cr}
\makeatother

\def\beq{\begin{equation}}
\def\eeq{\end{equation}}
\def\bea{\begin{eqnarray}}
\def\eea{\end{eqnarray}}
\def\be{\ba}
\def\ee{\ea}
\def\ba{\beq\new\begin{array}{c}}
\def\ea{\end{array}\eeq}
\def\f{1\over }
\def\nn{\nonumber}

\def\theequation{\arabic{equation}} 

\begin{document}

\begin{titlepage}
\setcounter{footnote}0
\begin{center}
\hfill UUITP-10/93\\
\hfill FIAN/TD-07/93\\
\hfill ITEP-M4/93
\footnote
{Preliminary version of this report was circulating as preprint
FIAN/TD-01/93,\ \ ITEP-M12/92}
\begin{flushright}{April 1993}\end{flushright}
\vspace{0.1in}{\LARGE\bf Generalized Kazakov-Migdal-Kontsevich Model:
group theory aspects}
\\[.4in]
{\Large  S.Kharchev\footnote{E-mail address:
tdparticle@glas.apc.org, serg@grotte.teorfys.uu.se},
A.Marshakov\footnote{E-mail address:
tdparticle@glas.apc.org, marshakov@nbivax.nbi.dk},
A.Mironov\footnote{E-mail address:
tdparticle@glas.apc.org, mironov@grotte.teorfys.uu.se}}\\
\bigskip {\it Institute for Theoretical Physics,
Uppsala University,  Box 803, S-751 08 Uppsala, Sweden\\
and\\ Theory Department ,  P.N.Lebedev Physics
Institute , Leninsky prospect, 53, Moscow,~117~924, Russia},
\bigskip
\\
{\Large A.Morozov\footnote{E-mail address:
tdparticle@glas.apc.org, morozov@vxdesy.desy.de}}\\
\bigskip {\it ITEP , Bol.Cheremushkinskaya, 25, Moscow, 117 259, Russia,\\ and
\\Institute for Theoretical Physics,
Uppsala University, Box 803, S-751 08 Uppsala, Sweden}
\end{center}
\bigskip

\newpage

\setcounter{page}2
\vspace{2.5in}
\centerline{\bf ABSTRACT}
\begin{quotation}
\noindent
The Kazakov-Migdal model, if considered as a functional of external fields,
can be always represented as an expansion over characters of $GL$ group.
The integration over "matter fields" can be interpreted as going over the
{\it model} (the space of all highest weight representations) of $GL$. In the
case of compact unitary groups the integrals should be substituted by {\it
discrete} sums over
weight lattice.  The $D=0$ version of the model is the Generalized
Kontsevich integral, which in the above-mentioned unitary (discrete)
situation coincides with partition function of the $2d$ Yang-Mills
theory with the target space of genus $g=0$ and $m=0,1,2$ holes. This
particular
quantity is always a bilinear combination of characters and appears to be
a Toda-lattice
$\tau$-function.  (This is generalization of the classical
statement that individual $GL$ characters are always singular KP
$\tau$-functions.) The corresponding element of the Universal Grassmannian
is very simple and somewhat similar to the one, arising in investigations
of the $c=1$ string models.  However, under certain circumstances the
formal sum over representations should be evaluated by steepest descent
method and this procedure leads to some more complicated elements of
Grassmannian.  This "Kontsevich phase" as opposed to the simple
"character phase" deserves further investigation.
\end{quotation}

\end{titlepage}
\clearpage
\newpage
\tableofcontents
\newpage
\setcounter{page}3
\setcounter{footnote}0
\section{Introduction}

Matrix models remain to be an an important part of modern string theory.
Originally developed as an approximate method mostly for the study of
Yang-Mills
theories, it appeared capable to extract some {\it exact} combinatorial
information from a
wider class of physical models. Restricted class of {\it eigenvalue} matrix
models provides now the most profound information about topology of module
spaces - the underlying structure of (topological) $2d$ gravity.
) $2d$ gravity.
Further progress of string theory
will depend on the possibility to apply  the methods, developed for
investigation of topological theories to generic string models. Within the
field of matrix models the problem is to find adequate generalization of
results, obtained for eigenvalue matrix models to the case of generic
matrix models, in particular with angular (unitary-matrix) degrees of freedom.
The main
puzzle is the adequate generalization of {\it integrable structure} of matrix
models, which plays the crucial role in the eigenvalue model case
\cite{Douglas,GMMMO,KMMMZ} (see also \cite{Mar93,Mor93} for a review).

At the boundary between eigenvalue and generic matrix models lies the so
called Kazakov-Migdal model \cite{KazMig}, which still can be explicitly
represented in terms of eigenvalues, but this representation is already
too complicated to be of use by itself. The model has three kinds of matrix
variables: Hermitean $N\times N$ matrices $\Phi^{(\alpha)}$ and
$L^{(\mu)}$ and unitary $N\times N$ matrices $U^{(\alpha\beta)}$,
$U^{(\alpha\mu)}$, and partition function is
\beq
{\cal Z}_{V}\{L\} = {\cal C}_{V}\{L\}
\int d\Phi^{(\alpha)} [dU^{(\alpha\beta)}] [dU^{(\alpha\mu)}]
\exp {\rm tr}\left(\sum V_\alpha(\Phi^{(\alpha)}) +\right.
$$
$$
 \left.+ \sum c_{\alpha\beta} \Phi^{(\alpha)} U^{(\alpha\beta)}
\Phi^{(\beta)} {U^{(\alpha\beta)}}^\dagger +
\sum c_{\alpha\mu} \Phi^{(\alpha)} U^{(\alpha\mu)}L^{(\mu)}
{U^{(\alpha\mu)}}^\dagger \right)
\label{g2km}
\eeq
In the original Kazakov-Migdal model $\alpha$ labels the points of regular
$D$-dimensional lattice, and $c_{\alpha\beta}\neq 0$ only for the closest
neighbours on the lattice. Also usually all the potentials $V_\alpha$ and
non-vanishing link couplings $c_{\alpha\beta}$ are the same, and
background fields $L^{(\mu)}$ are not introduced. A very special case of
the model (\ref{g2km}) is the Generalized Kontsevich Model \cite{KMMMZ}
with partition function
\be
{\cal Z}_V\{L\} = {\cal C}_V\{L\} \int d\Phi e^{-{\rm tr} V(\Phi) + {\rm
tr} \Phi ULU^\dagger}.
\label{gkm}
\ee
Below, we will use $G(KM)^2$ to denote the "generalized
Kazakov-\-Migdal-\-Kontsevich Model" (\ref{g2km}).

The crucial feature of $G(KM)^2$, which distinguishes it from the naive
multimatrix model
\beq
{\cal Z}_{V}^{mm}\{L\} = {\cal C}_{V}\{L\}
\int d\Phi^{(\alpha)}
\exp {\rm tr}\left(\sum V_\alpha(\Phi^{(\alpha)}) +\right.
$$
$$
 \left.+ \sum c_{\alpha\beta} \Phi^{(\alpha)}\Phi^{(\beta)}  +
\sum c_{\alpha\mu}\Phi^{(\alpha)}L^{(\mu)} \right)
\label{multmamo}
\eeq
is the possibility to represent it as an eigenvalue
model, i.e.  in the form of integral over eigenvalues $\phi_i^{(\alpha)}$,
$i = 1\ldots N$, of matrices $\Phi^{(\alpha)}$. In the case of the naive
multimatrix model it is only possible unless there are no closed loops
in ''target" space.
The central ingredient of
this transition is the Harish-Chandra-Itzykson-Zuber formula \cite{IZ}
\beq
I\{X,L\} =
\int [dU]_{n\times n} e^{{\rm tr} XULU^\dagger} =
\frac{(2\pi)^{n(n-1)\over 2}}{n!}
{{\det _{ij} e^{l_ix_j}}\over {\Delta(L)\Delta(X)}}
\label{IZ}
\eeq
Making use of this formula one can integrate over {\it link} angle variables
and rewrite (\ref{g2km}) as:
\beq
{\cal Z}_{V}\{L\} = {\cal C}_{V}\{L\}
\int \prod d\phi_i^{(\alpha)} \times
$$
$$
\times \exp \left(\sum V_\alpha(\phi_i^{(\alpha)} \right)
\prod I\{\Phi^{(\alpha)}, \Phi^{(\beta)};c_{\alpha\beta}\}
\prod I\{\Phi^{(\alpha)},L^{(\mu)};c_{\alpha\mu}\}
\label{g2km1}
\eeq
This is, however, not a very illuminating expression and some extra
ideas are required to  investigate its mathematical structure.

One of such
ideas is to make use of the group theory structure, implicit in
formula (\ref{g2km1}) and in the entire model (\ref{g2km}). In its turn, this
kind of structure should be related to integrability. We describe below
only some very simple results in this direction, concerning some
oversimplified versions of $G(KM)^2$, closely related to the
model (\ref{gkm}). Actually they are not very easy to generalize,
but this certainly deserves further investigation. The main motivation is
that Generalized Kontsevich model (\ref{gkm}) is not by itself sufficient
to exhaustively describe even generic $(p,q)$-string models with
$d\leq 2$ ($c\leq 1$) \cite{KM,Mir}: only  models of the $(p,1)$-type (and
their
topological deformations) are represented well enough in this
form \cite{LGGKM,Losev}. The necessary generalizations (related to
"multi-component" integable hierarchies) seem to require many background
fiels $L$ instead of a single one in (\ref{gkm}), and these can be
introduced  in the manner suggested by eq.(\ref{g2km}) without spoiling
the solvability of the model. The search of a reasonable generalization of
GKM in the framework of $G(KM)^2$ is an interesting problem for  future
investigation. The subject is also closely related to the popular
$2d$ Yang-Mills theory \cite{Mig}--\cite{CAMP}.

 The group-theoretical description of $G(KM)^2$ appears
to be somewhat ``orthogonal'' to the usual presentation of GKM
in existing literature.
In fact the models possess two different limits to be refered to as the
``character'' and ``Kontsevich'' phases.
The proper time-variables are different in two limits: they are
expanded in positive and negative powers of external fields $L$
in the character and Kontsevich phases respectively, so that the relation
between them is a kind of duality (``modular'') transformation
$L \longrightarrow 1/L$ ($\log G \longrightarrow 1/\log G$).
Remarkably, integrability occurs in both phases, but the detailed
structure is somewhat different: while it is KP pattern, that arises
in the simplest examples of Kontsevich phase,
in the character phase it is substituted
by a less specific Toda structure, which includes non-trivial
dependence on the zero-time (i.e. on the size of the matrix in the
matrix-model language).
This paper deals mostly with description of the character phase,
which, though much simpler, did not yet attract enough attention
in the theory of GKM. It is in this phase that generalization of GKM to
$G(KM)^2$ looks most straightforward.

\section{IZ formula and characters}

The proper derivation of eq.(\ref{IZ}) should be on the base of the DH
theorem \cite{DH} (see \cite{KMSW} for more comments on this
approach). Technically simpler is the derivation, proposed in
ref.\cite{Migd}, which is based on the equations
\ba
{\rm tr}\left(\frac{\partial}{\partial L_{tr}}\right)^n I\{X,L\} =
\left({\rm tr} X^n \right)I\{X,L\}; \\{\rm tr}\left(\frac{\partial}{\partial
X_{tr}}\right)^n I\{X,L\} =
\left({\rm tr} L^n \right)I\{X,L\}
\ea
together with the statement that $I\{X,L\}$ depends only on the
eigenvalues of $X$ and $L$. One can easily recognize in these equations
the standard differential equations for characters of $GL$ group:
\be
\Delta_n(G) \chi_R(G) = C_n(R) \chi_R(G),
\label{Caseqchar}
\ee
where $\chi_R(G) = {\rm Tr}_R G$ is the trace of group element $G$ in
representation $R$ and $\Delta_n$ is invariant differential operator of the
$n$-th order on group (for example, $\Delta_2$ is the Laplace operator on
group). Indeed,
\be
I\{X,L\} \equiv \frac{\hat \chi_R(G)}{d_R}
\sim \frac{\chi_R(G)}{d_R},
\label{IZverchar}
\ee
where $G = e^L$ and $X$ labels representation $R$ of dimension $d_R$,
while coefficient of proportionality between ${\hat \chi}_R$ and $\chi_R$ is
the ratio of Van-der-Monde
operators $\frac{\Delta(G)}{\Delta(L)} = \frac{\Delta(e^L)}{\Delta(L)} =
\prod_{i>j} \frac{e^{l_i}-e^{l_j}}{l_i-l_j}$.

In order to get rid of this ratio one can use a more complicated
matrix model (quantum mechanics on the coadjoint orbit) \cite{AFS}
\footnote{This formula can be considered as one-dimensional Wess-Zumino model.
Similarly, the two-dimensional $WZW$ model gives rise to Kac-Moody
characters.}:
\be\label{qmech}
\frac{\chi_R(e^L)}{d_R} \equiv \int [DU(t)] \exp \int {dt}{\rm tr}
X\left(U^{-1}(t)\partial_tU(t) + U^{-1}(t) LU(t)\right).
\ee
Perhaps, this 1-dimensional functional integral should be substituted
instead of
$$
{{\hat \chi}_R (e^L) \over d_R} = \int [dU] \exp {\rm tr} X U^{-1}LU^{\dagger}
$$
into eq.(\ref{g2km}) in
order to describe couplings to background fields\footnote{Note that
this procedure actually shifts the dimension $D \longrightarrow D+1$.}.

Eqs.(\ref{IZverchar}) and (\ref{IZ}) together are nothing but the well-known
Weyl character formula. Indeed, each irreducible representation $R$ of GL(N)
is uniquely
determined by the set of non-increasing positive numbers ("signature")
\beq\label{1}
R = (m_{1}, m_{2}, \ldots m_N), \ \ m_{1} \geq m_{2} \geq
\ldots \geq m_{N}
\eeq
where $m_{i}$ are eigenvalues of the highest vector:

\beq\label{2}
E_{ii}v = m_{i}v,\ \  E_{ij}v = 0, i<j .
\eeq
where $E_{ij}$ are representations of the GL(N) basis $e_{ij}$,
$(e_{ij})_{\alpha \beta} = \delta_{i \alpha } \delta_{j \beta }$ and

\beq\label{3}
[E_{ij}, E_{kl}] = \delta_{kj} E_{il} - \delta_{il} E_{kj}
\eeq

Every irreducible representation can be determined by the only
knowledge of the characters. First Weyl formula for (primitive)
characters reads:
\beq\label{4}
\chi = \frac{\det \lambda_{j}^{m_{i} + N - j}}{\prod_{i<j}(\lambda_{i} -
\lambda_{j})}
\eeq
where $\lambda_{i}$ are eigenvalues of the given matrix $G \in GL(N)$
and $\{ m_i \}$ being the lengths of the rows of corresponding Young table.

Equivalently, every irreducible representation can be characterized
by numbers
\beq\label{5}
(k_{1}, k_{2}, \ldots , k_{N}) ; k_{1} > k_{2} > \ldots k_{N} \geq 0
\eeq
where
\beq\label{6}
k_{i} = m_{i} + N - i
\eeq

The second Weyl formula for (primitive) characters (we will use also
$\Delta (\lambda) \equiv \prod_{i>j}(\lambda_{i} - \lambda_{j}$)) :

\beq\label{7}
\frac{\det \lambda_{j}^{k_{i}}}{\Delta (\lambda)} = \det P_{k_{i}-j+1}(T)
\eeq
where $P_{k}(T)$ are the Shur polynomials and times $\{ T_k \}$ appear in the
form of Miwa transformation

\beq\label{8}
T_{k} = \frac{1}{k} \sum_{i=1}^{N} \lambda_{i}^{k}
\eeq

It is only for integer values of $k_i$'s that characters indeed depend on
the {\it compact-group} elements $G = e^L$ rather than on those of the
non-compact algebra. In this case the value of the character at $G=I$
($L=0$) is equal to dimension $d_R = \chi_R(I)$ of the representation $R$,
\be\label{dimrep}
d_R = \Delta(k) = \prod_{i>j} {k_i - k_j\over i-j}.
\ee

Thus we obtain interpretation of the $G(KM)^2$ as a theory where integration
(sum) at every point $\alpha$ is going over all representations of $GL$
algebra, labeled by eigenvalues of matrices $\Phi_\alpha$ (i.e. integral
at every point $\alpha$ is over a {\it model} space of $GL$). This implies
a natural modification of the model for the case of unitary group:
integrals over eigenvalues of $\Phi_\alpha$ should be substituted by
discrete sums over module spaces, namely over integer-valued eigenvalues.
In the case of the original GKM (\ref{gkm}) this trick would give:
\be\label{gkmvev}
\int \Delta(x) ...   \rightarrow \sum d_R \chi_R(G) ...
\ee
If $n$ background fields are attached to the single point (to form a sphere
with $n$ holes), we have:
\be\label{ngkmvev}
\int (\Delta(x))^{(2-n)} ... \rightarrow
\sum_R d_R^{(2-n)} \prod \chi_R(G_\mu) ...
\ee

\section{Vertex operators and sewing (topology change)}

One can consider the terms with external fields in (\ref{g2km}) as insertions
of a kind of "vertex operators"
\be
{\cal V}_L(X) = \int [dU] e^{{\rm tr}LUXU^\dagger}
\label{vop1}
\ee
This expression is similar to $e^{ipx}$ in bosonic string model, and $L$ can
be considered as a kind of external momentum. Perhaps, this set of operators
should be enlarged to include more complicated integrals with $U$-dependent
measures, like\footnote{
See \cite{KMSW,Mor,Shat} for discussion of generalizations of
Itzykson-Zuber formula, adequate for calculations with such operators.}
\be
{\cal V}_L\{F,X\} = \int [dU] F(U) e^{{\rm tr}LUXU^\dagger}
\label{vop2}
\ee
as well as some integrals depending on $X = \Phi^{(\alpha)}$ at different sites
$\alpha$ (i.e. analogues of $h(\partial x)e^{ipx}$; also note that the
analogue of $\int e^{ipx(z)}d^2z$ for many site $G(KM)^2$ is rather
$\sum_\alpha {\cal V}_L(X^{(\alpha)})$ than ${\cal V}_L(X)$ itself).

Given an adequate set of vertex operators one can perform sewing procedure,
i.e. go from one collection of sites $\alpha$ and links
$\left.<\alpha\beta>\right|_{c_{\alpha\beta}\neq 0}$ to another
(topology changing). In the case of (\ref{g2km}) one can take two external
fields $L_\mu$ and $L_\nu$, attached to two sites $\alpha$ and $\beta$ and
integrate over these $L$'s with invariant measure
$\int\int dL_\mu dL_\nu \delta(L_\mu + L_\nu)$. As result the two fields
$X_\alpha$ and $X_\beta$ will be identified and insertion of

\beq\label{sewing}
\Delta(X^{(\alpha)})^{-2}\delta(X^{(\alpha)} - X^{(\beta)})
\eeq
appears in the integrand in (\ref{g2km}). Thus with this set of vertex
operators we obtain a specific sewing prescription. Still it seems to be
of interest, especially because of its relevance to the $2d$ Yang-Mills
theory, where sewing is known to produce a factor of $d_R^{-2}$. Sewing
procedure can be modified by changing (enlarging) the set of vertex operators
(intermediate states) or varying the form of the three-vertex. Note that so
defined sewing procedure is not the same as changing $c_{\alpha\beta}$  to
ours in the $G(KM)^2$ (1).

If vertex operators were introduced by the rule (\ref{qmech}) instead of
(\ref{vop1}),
\be\label{vert}
\hat V_L(X) = \int [dU(t)]  \exp \int {dt}{\rm tr}
X\left(U^{-1}(t)\partial_tU(t) + U^{-1}(t) LU(t)\right),
\ee
one should sew with the help of the measure $[d e^L]$ instead of $dL$.
In terms of characters the two relevant orthogonality conditions are:
\be\label{ortchar}
\int \hat\chi_{R_1}(e^L) \hat\chi_{R_2}(e^{-L}) dL = \delta_{R_1 R_2},
\nn \\
\int \chi_{R_1}(e^L) \chi_{R_2}(e^{-L}) [d e^L] = \delta_{R_1 R_2}.
\ee

One can consider the above described sewing procedure also more formally just
as a
consistent set of rules. Then, having the propagator (corresponding to a
sphere with two holes)
$$
\sum_\alpha \chi_\alpha (U_1) \chi_\alpha (U_2)
$$
and the wave function (corresponding to the sphere with a hole) $\sum_\alpha
\chi_\alpha (U) d_{\alpha}$, as well as the consistent rules of gluing
the hole by putting corresponsding $U=1$ and of gluing two holes just by
integrating with the proper measure (induced by (\ref{ortchar})), one still has
a freedom to choose three point function.
$$\sum_{\alpha,\beta,\gamma}
C_{\alpha\beta\gamma}\chi_\alpha\chi_\beta\chi_\gamma.
$$
Certainly, there
are some consistency requirements like duality. We do not know all possible
solutions of these, but two natural choices are (properly normalized)
delta-function coefficients and inverse Clebsh-Gordon coefficients. The
normalization might be fixed, say, from the requirement
$$
\sum_{\alpha,\beta} C_{\alpha\beta\gamma}d_{\alpha} d_ \beta = d_
\gamma.
$$
Then, the first case corresponds just to $2d$ Yang-Mills theory
\cite{Gross,Mig,Wit},
while the second one describes the correlators in $k\to\infty$ WZW theory
\cite{MS}.

Further point to be mentioned is that the quantities which we discussed
appear related to $2d$ Yang-Mills correlators with
fixed monodromies at the boundaries.
Og great interest are also
correlators of the Wilson loops. These objects are much more complicated,
even for the three-point function at genus zero the answer being
proportional to a Clebsh-Gordon coefficient. In general case, it is
represented in the form of group integrals and explicit answers are not
known.\footnote{We are grateful to
A.Gorsky and N.Nekrasov for the discussion of these points.}

Now let us brieflt discuss "illustrated version" of the sewing considered
in the section.
$G(KM)^2$ as defined by eq.(\ref{g2km}) allows a simple pictorial
representation.
``Feynman diagramms'' are
collections of points and crosses
(crosses standing for background fields), connected by links (which
depict IZ integrals (\ref{IZ})), see Fig.1a.
\iflandscape
\begin{figure}
\begin{picture}(40.00,90.00)(00.00,3.00)
\unitlength=0.80mm
\thicklines
\put(10.00,80.00){\line(1,0){20.00}}
\put(10.00,80.00){\line(1,2){5.00}}
\put(15.00,90.00){\makebox(0,0)[cc]{{\bf x}}}
\put(10.00,80.00){\circle*{2.00}}
\put(30.00,80.00){\circle*{2.00}}
\put(10.00,80.00){\line(-1,2){5.00}}
\put(5.00,90.00){\makebox(0,0)[cc]{{\bf x}}}
\put(10.00,75.00){\makebox(0,0)[cc]{$\alpha$}}
\put(33.00,77.00){\makebox(0,0)[cc]{$\beta$}}
\put(30.00,80.00){\line(1,2){5.00}}
\put(35.00,90.00){\makebox(0,0)[cc]{{\bf x}}}
\put(30.00,80.00){\line(0,-1){20.00}}
\put(30.00,60.00){\circle*{2.00}}
\put(33.00,63.00){\makebox(0,0)[cc]{$\gamma$}}
\put(30.00,60.00){\line(-1,-2){5.00}}
\put(30.00,60.00){\line(1,-2){5.00}}
\put(35.00,50.00){\makebox(0,0)[cc]{{\bf x}}}
\put(25.00,50.00){\makebox(0,0)[cc]{{\bf x}}}
\put(20.00,35.00){\makebox(0,0)[cc]{{a)}}}
\end{picture}
\begin{picture}(120.00,190.00)(-80.00,-30.00)
\unitlength=0.80mm
\thicklines
\put(30.00,50.00){\circle*{2.00}}
\put(30.00,50.00){\line(1,1){10.00}}
\put(40.00,60.00){\makebox(0,0)[cc]{{\bf x}}}
\put(40.00,40.00){\makebox(0,0)[cc]{{\bf x}}}
\put(20.00,60.00){\makebox(0,0)[cc]{{\bf x}}}
\put(20.00,40.00){\makebox(0,0)[cc]{{\bf x}}}
\put(30.00,50.00){\line(1,-1){10.00}}
\put(30.00,50.00){\line(-1,-1){10.00}}
\put(30.00,50.00){\line(-1,1){10.00}}
\put(30.00,45.00){\makebox(0,0)[cc]{$\alpha$}}
\put(30.00,20.00){\makebox(0,0)[cc]{{b)}}}
\put(30.00,00.00){\makebox(0,0)[cc]{Figure 1}}
\end{picture}
\begin{picture}(120.00,190.00)(-90.00,-30.00)
\unitlength=0.80mm
\thicklines
\put(20.00,50.00){\circle*{2.00}}
\put(20.00,50.00){\line(1,0){10.00}}
\put(30.00,50.00){\makebox(0,0)[cc]{{\bf x}}}
\put(20.00,45.00){\makebox(0,0)[cc]{$\alpha$}}
\put(20.00,20.00){\makebox(0,0)[cc]{c)}}
\end{picture}
\end{figure}
\fi

\ifportrait
\begin{figure}
\begin{picture}(40.00,90.00)(00.00,3.00)
\unitlength=0.80mm
\thicklines
\put(10.00,80.00){\line(1,0){20.00}}
\put(10.00,80.00){\line(1,2){5.00}}
\put(15.00,90.00){\makebox(0,0)[cc]{{\bf x}}}
\put(10.00,80.00){\circle*{2.00}}
\put(30.00,80.00){\circle*{2.00}}
\put(10.00,80.00){\line(-1,2){5.00}}
\put(5.00,90.00){\makebox(0,0)[cc]{{\bf x}}}
\put(10.00,75.00){\makebox(0,0)[cc]{$\alpha$}}
\put(33.00,77.00){\makebox(0,0)[cc]{$\beta$}}
\put(30.00,80.00){\line(1,2){5.00}}
\put(35.00,90.00){\makebox(0,0)[cc]{{\bf x}}}
\put(30.00,80.00){\line(0,-1){20.00}}
\put(30.00,60.00){\circle*{2.00}}
\put(33.00,63.00){\makebox(0,0)[cc]{$\gamma$}}
\put(30.00,60.00){\line(-1,-2){5.00}}
\put(30.00,60.00){\line(1,-2){5.00}}
\put(35.00,50.00){\makebox(0,0)[cc]{{\bf x}}}
\put(25.00,50.00){\makebox(0,0)[cc]{{\bf x}}}
\put(20.00,35.00){\makebox(0,0)[cc]{{a)}}}
\end{picture}
\begin{picture}(120.00,190.00)(-80.00,-30.00)
\unitlength=0.80mm
\thicklines
\put(30.00,50.00){\circle*{2.00}}
\put(30.00,50.00){\line(1,1){10.00}}
\put(40.00,60.00){\makebox(0,0)[cc]{{\bf x}}}
\put(40.00,40.00){\makebox(0,0)[cc]{{\bf x}}}
\put(20.00,60.00){\makebox(0,0)[cc]{{\bf x}}}
\put(20.00,40.00){\makebox(0,0)[cc]{{\bf x}}}
\put(30.00,50.00){\line(1,-1){10.00}}
\put(30.00,50.00){\line(-1,-1){10.00}}
\put(30.00,50.00){\line(-1,1){10.00}}
\put(30.00,45.00){\makebox(0,0)[cc]{$\alpha$}}
\put(30.00,20.00){\makebox(0,0)[cc]{{b)}}}
\put(30.00,00.00){\makebox(0,0)[cc]{Figure 1}}
\end{picture}
\begin{picture}(120.00,190.00)(-90.00,-30.00)
\unitlength=0.80mm
\thicklines
\put(20.00,50.00){\circle*{2.00}}
\put(20.00,50.00){\line(1,0){10.00}}
\put(30.00,50.00){\makebox(0,0)[cc]{{\bf x}}}
\put(20.00,45.00){\makebox(0,0)[cc]{$\alpha$}}
\put(20.00,20.00){\makebox(0,0)[cc]{c)}}
\end{picture}
\begin{picture}(120.00,190.00)(-90.00,16.00)
\unitlength=0.80mm
\thicklines
\put(19.00,70.00){\line(-3,1){20.00}}
\put(20.50,71.50){\line(-1,1){16.50}}
\put(20.50,71.50){\line(1,1){16.50}}
\put(22.00,70.00){\line(3,1){20.00}}
\put(0.75,82.75){\oval(9.50,12.50)}
\put(40.25,82.75){\oval(9.50,12.50)}
\put(19.00,70.00){\line(-3,-1){20.00}}
\put(20.50,68.50){\line(-1,-1){16.50}}
\put(20.50,68.50){\line(1,-1){16.50}}
\put(22.00,70.00){\line(3,-1){20.00}}
\put(0.75,57.25){\oval(9.50,12.50)}
\put(40.25,57.25){\oval(9.50,12.50)}
\put(19.50,19.00){\makebox(0,0){Figure2}}
\end{picture}
\end{figure}
\fi
A link appears between
the points $\alpha$ and $\beta$ whenever $c_{\alpha\beta} \neq 0$ and
the cross $\mu$ is attached to the point $\alpha$ whenever
$c_{\alpha\mu}\neq 0$. The simplest example (which we shall often refer
to as $D=0$ model) occurs when there is just a single point, Fig.1b.
Conventional GKM arises when the cross is also unique, Fig.1c.
If external fields are introduced by the rule (\ref{qmech}) instead of
(\ref{IZ}), links are effectively substituted by tubes, thus
a sphere with holes, Fig.2, emerges instead of a hedgehog in Fig.1b.
Refleting the difference between points and holes,
the relevant vertex operators (\ref{vert}) are, of course, non-local.
There are two ways of sewing two points, say, $\alpha$ and $\gamma$
in Fig.1a: either they can be connected by a new link, Fig.3a, or
the two attached external fields can be identified and then integrated
over according to the rule (\ref{sewing}), Fig.3b.

\ifportrait
\begin{figure}
\begin{picture}(120.00,190.00)(00.00,21.00)
\unitlength=0.80mm
\thicklines
\put(10.00,80.00){\line(1,0){20.00}}
\put(10.00,80.00){\circle*{2.00}}
\put(30.00,80.00){\circle*{2.00}}
\put(10.00,80.00){\line(-1,2){5.00}}
\put(10.00,75.00){\makebox(0,0)[cc]{$\alpha$}}
\put(33.00,77.00){\makebox(0,0)[cc]{$\beta$}}
\put(30.00,80.00){\line(1,2){5.00}}
\put(30.00,80.00){\line(0,-1){20.00}}
\put(30.00,60.00){\circle*{2.00}}
\put(33.00,63.00){\makebox(0,0)[cc]{$\gamma$}}
\put(30.00,60.00){\line(1,-2){5.00}}
\put(10.00,80.00){\line(1,-1){20.00}}
\put(15.00,90.00){\makebox(0,0)[cc]{{\bf x}}}
\put(5.00,90.00){\makebox(0,0)[cc]{{\bf x}}}
\put(35.00,90.00){\makebox(0,0)[cc]{{\bf x}}}
\put(35.00,50.00){\makebox(0,0)[cc]{{\bf x}}}
\put(25.00,50.00){\makebox(0,0)[cc]{{\bf x}}}
\put(20.00,33.00){\makebox(0,0)[cc]{a)}}
\put(10.00,80.00){\line(1,2){5.00}}
\put(30.00,60.00){\line(-1,-2){5.00}}
\end{picture}
\begin{picture}(120.00,190.00)(-30.00,21.00)
\unitlength=0.80mm
\thicklines
\put(-15.00,20.00){\makebox(0,0)[cc]{Figure 3}}
\put(5.00,80.00){\line(1,0){20.00}}
\put(5.00,80.00){\circle*{2.00}}
\put(25.00,80.00){\circle*{2.00}}
\put(5.00,80.00){\line(-1,2){5.00}}
\put(5.00,75.00){\makebox(0,0)[cc]{$\alpha$}}
\put(28.00,77.00){\makebox(0,0)[cc]{$\beta$}}
\put(25.00,80.00){\line(1,2){5.00}}
\put(25.00,80.00){\line(0,-1){20.00}}
\put(25.00,60.00){\circle*{2.00}}
\put(28.00,63.00){\makebox(0,0)[cc]{$\gamma$}}
\put(25.00,60.00){\line(1,-2){5.00}}
\put(5.00,80.00){\line(1,-1){20.00}}
\put(0.00,90.00){\makebox(0,0)[cc]{{\bf x}}}
\put(30.00,90.00){\makebox(0,0)[cc]{{\bf x}}}
\put(30.00,50.00){\makebox(0,0)[cc]{{\bf x}}}
\put(15.00,33.00){\makebox(0,0)[cc]{b)}}
\put(15.00,70.00){\makebox(0,0)[cc]{{\bf X}}}
\end{picture}
\begin{picture}(70.00,60.00)(-25.00,00.00)
\unitlength=0.80mm
\thicklines
\put(20.00,70.00){\oval(20.00,20.00)[lb]}
\put(14.00,60.00){\oval(12.00,20.00)[rt]}
\put(10.00,70.00){\line(3,0){6.00}}
\put(20.00,60.00){\circle*{2.00}}
\put(10.00,70.00){\makebox(0,0)[cc]{{\bf x}}}
\put(20.00,70.00){\oval(20.00,20.00)[rb]}
\put(26.00,60.00){\oval(12.00,20.00)[lt]}
\put(30.00,70.00){\line(-3,0){6.00}}
\put(30.00,70.00){\makebox(0,0)[cc]{{\bf x}}}
\put(20.00,60.00){\line(-1,-1){15.00}}
\put(20.00,60.00){\line(0,-1){20.00}}
\put(20.00,60.00){\line(1,-1){15.00}}
\put(23.00,63.00){\makebox(0,0)[cc]{$\alpha$}}
\put(20.00,23.00){\makebox(0,0)[cc]{a)}}
\put(5.00,45.00){\makebox(0,0)[cc]{{\bf x}}}
\put(35.00,45.00){\makebox(0,0)[cc]{{\bf x}}}
\put(20.00,40.00){\makebox(0,0)[cc]{{\bf x}}}
\end{picture}
\begin{picture}(70.00,60.00)(-75.00,00.00)
\unitlength=0.80mm
\thicklines
\put(20.00,70.00){\oval(20.00,20.00)[t]}
\put(16.00,70.00){\oval(6.00,4.00)[b]}
\put(24.00,70.00){\oval(6.00,4.00)[b]}
\put(16.00,68.00){\oval(3.00,2.00)[t]}
\put(16.00,70.00){\oval(12.00,20.00)[lb]}
\put(24.00,68.00){\oval(3.00,2.00)[t]}
\put(24.00,70.00){\oval(12.00,20.00)[rb]}
\put(16.00,60.00){\line(-3,-1){20.00}}
\put(19.00,60.00){\line(-1,-1){18.00}}
\put(19.00,60.00){\line(-1,-3){6.50}}
\put(21.00,60.00){\line(1,-3){6.50}}
\put(21.00,60.00){\line(1,-1){18.00}}
\put(24.00,60.00){\line(3,-1){20.00}}
\put(20.00,38.00){\oval(15.50,12.00)}
\put(-2.25,47.25){\oval(9.50,12.50)}
\put(42.25,47.25){\oval(9.50,12.50)}
\put(20.00,22.00){\makebox(0,0)[cc]{b)}}
\put(-10.00,10.00){\makebox(0,0){Figure 4}}
\end{picture}
\end{figure}
\fi

\iflandscape
\begin{figure}
\begin{picture}(70.00,10.00)
\unitlength=0.80mm
\thicklines
\put(19.00,70.00){\line(-3,1){20.00}}
\put(20.50,71.50){\line(-1,1){16.50}}
\put(20.50,71.50){\line(1,1){16.50}}
\put(22.00,70.00){\line(3,1){20.00}}
\put(0.75,82.75){\oval(9.50,12.50)}
\put(40.25,82.75){\oval(9.50,12.50)}
\put(19.00,70.00){\line(-3,-1){20.00}}
\put(20.50,68.50){\line(-1,-1){16.50}}
\put(20.50,68.50){\line(1,-1){16.50}}
\put(22.00,70.00){\line(3,-1){20.00}}
\put(0.75,57.25){\oval(9.50,12.50)}
\put(40.25,57.25){\oval(9.50,12.50)}
\put(19.50,20.00){\makebox(0,0){Figure2}}
\end{picture}
\begin{picture}(120.00,190.00)(-40.00,00.00)
\unitlength=0.80mm
\thicklines
\put(10.00,80.00){\line(1,0){20.00}}
\put(10.00,80.00){\circle*{2.00}}
\put(30.00,80.00){\circle*{2.00}}
\put(10.00,80.00){\line(-1,2){5.00}}
\put(10.00,75.00){\makebox(0,0)[cc]{$\alpha$}}
\put(33.00,77.00){\makebox(0,0)[cc]{$\beta$}}
\put(30.00,80.00){\line(1,2){5.00}}
\put(30.00,80.00){\line(0,-1){20.00}}
\put(30.00,60.00){\circle*{2.00}}
\put(33.00,63.00){\makebox(0,0)[cc]{$\gamma$}}
\put(30.00,60.00){\line(1,-2){5.00}}
\put(10.00,80.00){\line(1,-1){20.00}}
\put(15.00,90.00){\makebox(0,0)[cc]{{\bf x}}}
\put(5.00,90.00){\makebox(0,0)[cc]{{\bf x}}}
\put(35.00,90.00){\makebox(0,0)[cc]{{\bf x}}}
\put(35.00,50.00){\makebox(0,0)[cc]{{\bf x}}}
\put(25.00,50.00){\makebox(0,0)[cc]{{\bf x}}}
\put(20.00,30.00){\makebox(0,0)[cc]{a)}}
\put(10.00,80.00){\line(1,2){5.00}}
\put(30.00,60.00){\line(-1,-2){5.00}}
\end{picture}
\begin{picture}(120.00,190.00)(-30.00,00.00)
\unitlength=0.80mm
\thicklines
\put(-15.00,20.00){\makebox(0,0)[cc]{Figure 3}}
\put(5.00,80.00){\line(1,0){20.00}}
\put(5.00,80.00){\circle*{2.00}}
\put(25.00,80.00){\circle*{2.00}}
\put(5.00,80.00){\line(-1,2){5.00}}
\put(5.00,75.00){\makebox(0,0)[cc]{$\alpha$}}
\put(28.00,77.00){\makebox(0,0)[cc]{$\beta$}}
\put(25.00,80.00){\line(1,2){5.00}}
\put(25.00,80.00){\line(0,-1){20.00}}
\put(25.00,60.00){\circle*{2.00}}
\put(28.00,63.00){\makebox(0,0)[cc]{$\gamma$}}
\put(25.00,60.00){\line(1,-2){5.00}}
\put(5.00,80.00){\line(1,-1){20.00}}
\put(0.00,90.00){\makebox(0,0)[cc]{{\bf x}}}
\put(30.00,90.00){\makebox(0,0)[cc]{{\bf x}}}
\put(30.00,50.00){\makebox(0,0)[cc]{{\bf x}}}
\put(15.00,30.00){\makebox(0,0)[cc]{b)}}
\put(15.00,70.00){\makebox(0,0)[cc]{{\bf X}}}
\end{picture}
\begin{picture}(70.00,60.00)(00.00,100.00)
\unitlength=0.80mm
\thicklines
\put(20.00,70.00){\oval(20.00,20.00)[lb]}
\put(14.00,60.00){\oval(12.00,20.00)[rt]}
\put(10.00,70.00){\line(3,0){6.00}}
\put(20.00,60.00){\circle*{2.00}}
\put(10.00,70.00){\makebox(0,0)[cc]{{\bf x}}}
\put(20.00,70.00){\oval(20.00,20.00)[rb]}
\put(26.00,60.00){\oval(12.00,20.00)[lt]}
\put(30.00,70.00){\line(-3,0){6.00}}
\put(30.00,70.00){\makebox(0,0)[cc]{{\bf x}}}
\put(20.00,60.00){\line(-1,-1){15.00}}
\put(20.00,60.00){\line(0,-1){20.00}}
\put(20.00,60.00){\line(1,-1){15.00}}
\put(23.00,63.00){\makebox(0,0)[cc]{$\alpha$}}
\put(20.00,23.00){\makebox(0,0)[cc]{a)}}
\put(5.00,45.00){\makebox(0,0)[cc]{{\bf x}}}
\put(35.00,45.00){\makebox(0,0)[cc]{{\bf x}}}
\put(20.00,40.00){\makebox(0,0)[cc]{{\bf x}}}
\end{picture}
\begin{picture}(70.00,60.00)(-80.00,100.00)
\unitlength=0.80mm
\thicklines
\put(20.00,70.00){\oval(20.00,20.00)[t]}
\put(16.00,70.00){\oval(6.00,4.00)[b]}
\put(24.00,70.00){\oval(6.00,4.00)[b]}
\put(16.00,68.00){\oval(3.00,2.00)[t]}
\put(16.00,70.00){\oval(12.00,20.00)[lb]}
\put(24.00,68.00){\oval(3.00,2.00)[t]}
\put(24.00,70.00){\oval(12.00,20.00)[rb]}
\put(16.00,60.00){\line(-3,-1){20.00}}
\put(19.00,60.00){\line(-1,-1){18.00}}
\put(19.00,60.00){\line(-1,-3){6.50}}
\put(21.00,60.00){\line(1,-3){6.50}}
\put(21.00,60.00){\line(1,-1){18.00}}
\put(24.00,60.00){\line(3,-1){20.00}}
\put(20.00,38.00){\oval(15.50,12.00)}
\put(-2.25,47.25){\oval(9.50,12.50)}
\put(42.25,47.25){\oval(9.50,12.50)}
\put(20.00,22.00){\makebox(0,0)[cc]{b)}}
\put(-10.00,10.00){\makebox(0,0){Figure 4}}
\end{picture}
\end{figure}
\fi

When the latter prescription
is used, the $D=0$ model acquires contributions like Fig.4a or Fig.4b,
depending on the way the external fields are introduced.
These
pictures makes identification of the $D=0\ $ $G(KM)^2$ with $2d$ YM theory
rather appealing. They also suggest that appropriate generalization of
IZ integrals, which would effectively turn links into tubes, deserves
being looked for. The last task is, however, beyond the scope of this
paper.

\section{Determinant representations of $\tau$-functions}

In what follows we are going to discuss integrability properties of partition
functions, constructed from characters. Therefore we need to
remind briefly the main notions from the theory of $\tau$-functions (more
formulas which will be used in the next sections are contained in the
Appenidices A).

We restrict ourselves to consideration of ordinary KP and Toda-lattice
$\tau$-functions, which can be constructed as correlators in the theory of
free fermions $\psi, \psi^{\ast}$ ($b,c$-system of spin $1/2$). The basic
quantity is the ratio of fermionic correlators,
\be\label{tau}
\tau_N(t,\bar t \mid {\cal G}) \equiv \frac
{\langle N \mid e^H {\cal G} e^{\bar H} \mid N \rangle}
{\langle N \mid  {\cal G}  \mid N \rangle},
\ee
in the theory of free 2-dimensional fermionic fields  $\psi (z)$,
$\psi^{\ast} (z)$  with the action $\int
\psi^{\ast} \bar \partial \psi $ (see the Appendix A1 for the definitions
and notations). In the definition (\ref{tau})
\beq\label{Ham}
H = \sum_{k > 0} t_kJ_k; \ \ \ \bar H = \sum_{k>0} \bar t_k J_{-k},
\eeq
and the currents are defined to be
\beq\label{cur}
J(z) = \psi^{\ast} (z)\psi (z)\hbox{; }
\eeq
where  $\psi (z) =
\sum _{\bf Z}
\psi _nz^n\ dz^{1/2} $ , $\psi^{\ast} (z) =
\sum _{\bf Z}
\psi^{\ast} _nz^{-n-1}\ dz^{1/2}$.

This expression for $\tau$-function is rather famous. Let us point out that
this
$\tau$-function is the $\tau$-function of the most general one-component
hierarchy
which is Toda lattice hierarchy. Putting all negative times to be zero as
well as
discrete index $n$, one obtains usual KP hierarchy. An element

\beq\label{elGr}
 {\cal G} =\
:\exp \{ \sum_{m,n} {\cal A}_{mn}\psi^{\ast} _m\psi _n\}:
\eeq
is an element of the group $GL(\infty)$ realized in the infinite dimensional
Grassmannian. The normal ordering should be understood here with respect to
any $|k\rangle$ vacuum,\footnote {Indeed, it is rather crucial point and we
return to possible choices of the normal orderings in the section 5.}
where vacuum states are defined by conditions

\beq\label{vac}
\psi _m|k\rangle  = 0\ \ m < k\hbox{ , }  \psi^{\ast} _m|k\rangle  = 0\ \
m \geq  k.
\eeq

After Miwa transformation of time variables,\footnote{
We choose number $n$ of all the four $\lambda$-variables the same. This
restriction can be take off eliminating $\lambda$ by puting it
equal to $\infty$, while every $\bar\lambda$-variable is eliminated by
taking it equal to zero.
}

\be\label{Miwavar}
t_k = -\frac{1}{k} \sum_{i = 1}^n (
\mu_i^{-k} - \lambda_i^{-k}); \nn \\
\bar t_k = -\frac{1}{k}
\sum_{i = 1}^{n} ( \bar\mu_i^{k} - \bar\lambda_i^{k}),
\ee
the $\tau$-functions acquires the form of \footnote{The simplest way to
demonstrate this is to apply the bosonization formulas as in the Appendix A1.}

\be
\tau_N(t,\bar t \mid {\cal G}) =
{1\over \Delta (\lambda ) \Delta (\mu )
\Delta (\bar\lambda ) \Delta (\bar\mu)}\times\\
\times \frac {\langle N \mid
\prod_{i=1}^n \psi(\lambda_i)\prod_{i=1}^n\psi^{\ast}(\mu_i) {\cal G}
\prod_{i=1}^n\psi(\bar\mu_i)
\prod_{i=1}^n\psi^{\ast}(\bar\lambda_i)
\mid N \rangle}
{\langle N \mid  {\cal G} \mid N \rangle}
\label{miwatau}
\ee
Now Wick theorem can be used  (the choice of ${\cal G}$ as quadratic exponent
in fermionic fields  is important here) in order to rewrite the r.h.s. in the
form of determinant of $2n\times 2n$ matrix, with {\it pair} correlators
as the entries:

\beq\label{gendet}
\tau_N(t,\bar t \mid {\cal G}) =
{1\over \Delta (\lambda ) \Delta (\mu )
\Delta (\bar\lambda ) \Delta (\bar\mu)}
{\rm det} \left(
\begin{array}{cc}
\frac {\langle N \mid
 \psi(\lambda_i)\psi^{\ast}(\mu_i) {\cal G}\mid N \rangle}
{\langle N \mid  {\cal G} \mid N \rangle}
&
\frac {\langle N \mid
 \psi(\lambda_i) {\cal G} \psi^{\ast}(\bar\lambda_i) \mid N \rangle}
{\langle N \mid  {\cal G} \mid N \rangle}\\
\frac {\langle N \mid
\psi^{\ast}(\mu_i) {\cal G} \psi(\bar\mu_i)
\mid N \rangle}
{\langle N \mid  {\cal G} \mid N \rangle}
  &
\frac {\langle N \mid
{\cal G} \psi^{\ast}(\bar\lambda_i) \psi(\bar\mu_i)
\mid N \rangle}
{\langle N \mid  {\cal G} \mid N \rangle}
\end{array} \right)
\label{det4cor}
\eeq

This expresion can be considerably simplified for special choices of
points $\lambda$. There are two especially important particular cases.
First, let us put all $\bar\lambda$ and $\bar\mu$ equal to
zero. Then the right lower block of the matrix in (\ref{det4cor})
is singular and the contributions of the off-diagonal blocks can be neglected.
The singularity is canceled by the zeroes, coming from the Van-der-Monde
determinants, and we obtain:

\be\label{9}
\tau_N(t \mid {\cal G})
= \frac{\prod (\mu_i-\lambda_j)\ \det_{ij}
C^{11}(\mu_i,\lambda_j)}
{\Delta(\lambda)\Delta(\mu)}
\ee
with
\beq\label{10}
C^{11}(\mu_i,\lambda_j) \equiv \frac {\langle N \mid
 \psi(\lambda_i)\psi^{\ast}(\mu_i) {\cal G}\mid N \rangle}{
\langle N \mid  {\cal G} \mid N \rangle}.
\eeq
One can further tend all the $\lambda$  to $\infty$ and obtain:

\be\label{detrep}
\tau_N(t \mid {\cal G})
= \frac{ {\rm det}_{ij}
\phi_i(\mu_j)}
{\Delta(\mu)}
\ee
with

\be\label{basvec}
\phi_i(\mu) \equiv {1 \over (i-1)!} \lim_{t \to 0} \partial ^{i-1}_t
\frac {\langle N \mid
\psi(t^{-1})\psi^{\ast}(\mu ) {\cal G}\mid N \rangle}
{\langle N \mid  {\cal G} \mid N \rangle}
= \mu^{i-1}( 1 + {\cal O}(\mu^{-1})).
\ee
This is the expression for KP $\tau$-function in Miwa
coordinates, discussed in much details in \cite{KMMMZ}.

Another particular case of eq.(\ref{det4cor}) arises when
\beq
{\rm all}\ \lambda \rightarrow \infty,\ \ \ \ \
{\rm while}\
{\rm all}\
\bar\lambda \rightarrow 0
\label{notilde}
\eeq
Then (\ref{det4cor}) turns into
not simple at all, and in fact it is better to take the
limit (\ref{notilde}) directly in the original correlator (\ref{miwatau}).
Then (\ref{miwatau}) turns into (see the Appendix A2)

\beq
\tau_N(t,\bar t \mid {\cal G}) =  \hbox{ Van-der-Mondes }\times
$$
$$
\times \frac {\langle N-n \mid
\prod_{i=1}^n \psi(\lambda_i) {\cal G} \prod_{i=1}^n\psi^{\ast}(\bar\lambda_i)
\mid N-n \rangle} {\langle N \mid  {\cal G} \mid N \rangle},
\label{miwataunotilde}
\eeq
and application of Wick theorem provides a determinant of $n\times n$
matrix:

\beq\label{Toda}
\new
\begin{array}{c}
\tau_N(t,\bar t) = \frac{\langle N-n|{\cal G}|N-n\rangle}{\Delta (z^{-1})
\Delta (w)} \times \nn\\
\times \det \frac{\langle N-n+1|e^{\sum \frac{1}{k}J_{k}z^{-k}_i} \;
{\cal G} \; e^{\sum \frac{1}{k}\bar
J_{k}w^{k}_j}|N-n+1\rangle}{\langle N-n|{\cal G}|N-n \rangle}
\end{array}\label{det2cornotilde}
\eeq

\bigskip

Among all above determinants of correlators, the representation
(\ref{basvec}) can be immediately interpreted in terms of Baker-Akhiezer
(BA) function. Indeed, it can be expressed through the fermionic correlator
as follows:

\be\label{BA}
\Psi_N(\mu,t_k) \equiv  \frac{\tau (t_k- {1 \over
k\mu^k})}{\tau (t_k)}e^{\sum t_k\mu^k} =
\frac{\langle N+1 \mid e^H \psi (\mu) {\cal G}
\mid N \rangle}{\langle N \mid e^H {\cal G} \mid N \rangle}.
\ee
The analogous expression is certainly correct for the conjugated function.
If we consider Toda lattice case, i.e. allows one to put fermions to the
right side with respect to ${\cal G}$, we get four different BA functions. Now
we easily express the entries of the determinant
(\ref{basvec}) through one of these functions:

\be\label{BAbasvec}
\Psi_N(\mu,t_k)\left|_{{t_1=x}\atop{t_k=0, k\ne 1}} = \sum_i
{\phi_i(\mu)x^i \over i!}\right.,
\ee
i.e.

\be\label{basvecBA}
\left.\phi_i(\mu) = \frac{\partial^{i-1} \Psi_N(\mu,t_k)}{\partial x ^{i-1}}
\right|_{{t_1=x}\atop{t_k=0, k\ne 1}}.
\ee
As for the Toda lattice case, the entries of the determinant (\ref{gendet})
are equal to the BA-functions only asymptotically. Put it differently, BA
functions describes the transition amplitudes to the vacuum state, while
the determinant (\ref{gendet}) contains the transition amplitudes between
all one-particle states.

\bigskip

\section{Characters of GL group as singular KP $\tau$-functions}

\subsection{A notion of singular $\tau$-function}

Characters of irreps of GL(N) and SL(N) (see \cite{Zhelobenko}) are
given by formulas (\ref{4}) and (\ref{7}). Now we are going to demonstrate that
these are
indeed $\tau$-functions. At first, we consider the $\tau$-function in terms
of times. We follow the classical paper \cite{DJKM} and consider for the sake
of brevity only KP, not Toda hierarchy.

There is a subspace in the Grassmannian which corresponds to the sigular
$\tau$-functions. It means that it can be described in the following equivalent
ways:

{\it i)} The v.e.v. of the element ${\cal G}$ is equal to zero.

{\it ii)} It can not be written in the canonical exponential form (\ref{elGr})
in some of natural normal orderings.

{\it iii)} It can be immediately related to the characters of group
$GL(N)$ at {\it finite} $N$.

{\it iv)} The $\tau$-function is a polynomial of times (rational solutions to
the KP hierarchy).

{\it v)} The $\tau$-function can be represented in the determinant form of
(\ref{detrep}), but no preserving the property of $\phi_i(\mu) \sim
\mu^{i-1}+O(1/\mu)$.

Now we are going to comment more on these statements and their interrelations.
To begin with, we discuss the case of {\it i)}. It implies that the expressions
of the type (\ref{basvec}) require some regularization which changes the
asymptotics
of the entries in the determinant and lead to the statement {\it v)}. We
consider how to do this in the subsection 5.3.

Let us stress that the exponential form of the element of the Grassmannian
(\ref{elGr}) hints to the non-zero v.e.v. Indeed, one should also take into
account
the normal ordering. There are two natural normal orderings - these are
with respect to the vacuum state $|0>$ and to the empty
vacuum state $|+\infty>$ (see the Appendices A1 and A3). The first one implies
that
all annihilation operators to be shifted to the right are $\psi_k^\ast$ with
$k\ge 0$ and $\psi_k$ with $k<0$. The second normal ordering implies all
operators with no asterisques standing at the right.

If we represent now the element of the Grassmannian as a normal ordered
exponent bilinear in fermions, then the matrices ${\cal A}_{mn}$ of the
bilinear forms (\ref{elGr}) for
different normal orderings and for the exponent unordered at all are related
in an explicit way \cite{DJKM}.
It is obviously that one can not obtain zero v.e.v. from non-singular
${\cal A}_{mn}^{unord}$, as it should be always 1. The same is correct for the
non-singular ${\cal A}_{mn}^{1ord}$. Thus, one is led to consider the second
normal ordering, which is singular in this case\footnote{Note that the
first ordering, associated to forced hierarchies \cite{KMMOZ,KMMM}
which describe matrix models before taking the continuum limit, leads to an
infinite matrix ${\cal G}_{mn}$ in (\ref{elGr}). In the case, the only
proper normal ordering of mentioned above is the first one.}. Thus, we
have explained the statement {\it ii)}.

\subsection{The characters and Shur polynomials}

Let us demonstrate the manifest construction of an element of the Grassmannian
corresponding to the singular $\tau$-functions. Hereafter, we consider only the
second normal ordering. Then, we consider ${\cal G}^{(m,n)}=\ :
\exp\{(\psi_{-m}+\psi_n)(\psi_{-m}^{\ast}
+\psi_n^{\ast})\} :$ ($m$ and $n$ are integers) which simplifies when acting on
the right vacuum:

\beq\label{singelGr}
{\cal G}^{(m,n)}|0>\  = [1+(\psi_{-m}+\psi_n)(\psi_{-m}^{\ast}+
\psi_n^{\ast})]|0> \ =  \psi_{-m}^{\ast}\psi_n|0>.
\eeq
Now one can trivially calculate the $\tau$-function (see (\ref{A17}),
(\ref{A18}), (\ref{A22}) and (\ref{A23})):

\beq
\tau({t_k},{\cal G}^{(m,n)}) =\  <0|e^{H(t)}\psi_{-m}^{\ast}\psi_n|0>\  =
$$
$$
=\  <0|\psi_{-m}^{\ast}(t)\psi_n(t)|0> \ = \sum_{l \ge 0} P_{l+m}(-t_k)
P_{n-l}(t_k),\label{stb}
\eeq
where Shur polynomials $P(t_k)$ are defined

\beq\label{Shur}
\exp\{\sum_{k>0}t_kx^k\} \equiv \sum _{k>0} P_k(t_k)x^k.
\eeq
Comparing this result with the second Weyl formula for (primitive) characters,
one can see that (\ref{stb}) describes the character $\chi_{m,n}$ of the
representation
corresponding to the "hook" Young diagramm with the row of the length $n+1$
and the column of the length $m$.\footnote{The simplest way to do this is
to calculate the matrix of rotations (\ref{A16}) for the element
${\cal G}^{(m,n)}$ and, then, to use formulas (\ref{A30}), (\ref{A31}).
It automatically produces the correct determinant form as in formula (\ref{7}).
One can also just use the identities for Shur polynomials obtaining the
same result.}

Now it is trivial to generalize the consideration. Indeed, let us consider more
complicated element of the Grassmannian ${\cal G}^{(\{m\},\{n\})} \equiv \ :
\exp\{\prod_i (\psi_{-m_i}+\psi_{n_i})(\psi_{-m_i}^{\ast}
+\psi_{n_i}^{\ast})\}: $ which is again equal on the right vacuum to

\beq\label{singelGrgen}
{\cal G}^{(\{m\},\{n\})}|0> \ = \prod_i \psi_{-m_i}^{\ast}\psi_{n_i} |0>
\eeq
and the $\tau$-function is

\beq
\tau({t_k},{\cal G}^{(\{m\},\{n\})}) = \ <0|e^{H(t)}\prod_i\psi_{-m_i}^{\ast}
\psi_{n_i}|0>\  =
$$
$$
\ <0|\prod_i\psi_{-m_i}^{\ast}(t)\psi_{n_i}(t)|0>\ = \det_{ij} <0|
\psi_{-m_i}^{\ast}(t)\psi_{n_j}(t)|0> \ = \det_{ij} \chi_{m_in_j}.
\label{hooks}
\eeq
Now, from the textbooks (see, for example, \cite{Ltw}) it is known that if one
chooses a sufficiently large integer $N$ and parameterize an element
${\cal G}$ of $GL(N)$ as $t_k \equiv {1 \over k} \hbox{tr} \ {\cal G}^k$,
then the character corresponding to the Young diagramm with $l$ "hooks" (i.e.
parameterized
by $n_i,m_i, \ \ \ i=1, \ldots , l$) is a polynomial in $t_k$ independent of
the choice of large $N$,\footnote{Which proves the statement {\it iv)}.}
which is given explicitly by formula (\ref{hooks}).
Again, it is also easy to turn this formula to the second Weyl formula (see
the last footnote).

Thus, we have demonstrated the statements {\it iii)} and {\it iv)} and are
going to explain how it is possible using Miwa parameterization, to reproduce
in the framework of the present paper the first Weyl formula. Simultaneously
we explain the statement {\it v)}.

\subsection{Singular $\tau$-functions in Miwa variables}

Thus, we derive the expression for basis vector for the case of the
singular $\tau$-functions.
To do this, let us consider again in details all the stuff connected
with the elementary representation  (of $GL(N)$-group) given by hook Young
diagramm which contains a row of $n+1$ boxes and a column of $m$ boxes.
We already established that the corresponding $\tau$-function is equal to
proper determinant with entries linear in Shur
polynomials (\ref{Shur}) and simultaneously to the following expression:

$$
\chi_{m,n} = \sum_{k=0}^n  P_{k+m}(\{-t_k\})P_{n-k}(\{t_k\}).
$$
When $m=1$ this is equal to $P_{n+1}(\{t_k\})$ and should be equal to the
character
of the representation given by the Young diagramm consisting of a line of $n$
boxes.
To convince oneself that all listed formulas are correct it is sufficient to
write down
the manifest expressions for Shur polynomials in Miwa variables (their number
will
be $N$, therefore all determinants have a size $N\times N$):

$$
P_n(t) = {(-)^n \over \Delta (\mu )} \det \left|
\begin{array}{c}
\{ \mu_i^{-n} \} \\ \{\mu_i\} \\ \ldots

\end{array}\right|
$$
There is an analogous expression for $P(-t)$ through Miwa variables:

$$
P_n(-t) = {(-)^n \over \Delta (\mu )} \det \left|
\begin{array}{c}
\{\mu_i^0 \} \\ \{\mu_i\} \\ \ldots \\ \{\mu_i^{n-1}\} \\ \{\mu_i^{-1}\} \\
\{\mu_i^{n}\} \\ \ldots
\end{array}\right|
$$
At last, one can write down the expression $\sum P(-t)P(t)$ again through Miwa
variables:

\beq\label{elchar}
\sum_{k=0}^{n} P_{k+m}(-t)P_{n-k}(t) = \sum_{k=1}^N {(-)^n \delta_{m,k}\over
\Delta (\mu )} \det \left|
\begin{array}{c}
\{\mu_i^{-n-1} \} \\ \{\mu_i^0\} \\ \ldots \\ \{\mu_i^{k-2}\} \\ \{\mu_i^k\} \\
\ldots
\end{array}\right|
\eeq
The last expression can be obtained from (\ref{4}) by substituition
$\lambda_i \to
\mu_i^{-1}$. Then, the dependence of $N$ can be got rid of and
the result is the determinant with the lines ${\mu_j^{-m_i+i-1}}$ such
that $j$ goes from 1 to infinity and $m_i\neq 0$ only for $i\ll \infty$. Thus,
we
produce all the expressions for characters like above. The last thing is still
remained
is to prove the last formula. We can do this in the standard framework of
\cite{KMMMZ} regularizing the expression for basis vector to have non-zero
vacuum
expectation value of the element of $GL(\infty)$. Namely, let us consider the
renormalized Clifford element ${\cal G}$ (let $m \geq 1;n\geq 0$):
\begin{equation}\label{regelGr}
{\cal G}^{(m,n)} = 1+(A_{1}\psi_{-m}^{\ast} +
A_{2}\psi_{n}^{\ast})(B_{1}\psi_{-m} +
B_{2}\psi_{n})
\end{equation}
such that
\begin{equation}\label{regelGrvac}
{\cal G}^{(m,n)}|0\rangle = (1+A_{2}B_{2} +
A_{1}B_{2}\psi_{-m}^{\ast}\psi_{n})|0\rangle
\end{equation}
and, for definiteness, let us choose
$
A_{1}B_{2} = -1, 1+A_{2}B_{2}=\epsilon .
$
Then basic vectors are defined in the standard manner (\ref{basvec})
\begin{equation}\label{basvecreg}
\phi_{i}^{(m,n)}=\left.\frac{1}{(i-1)!}\partial_{t}^{i-1}
\frac{\langle0|\exp\left[\sum\frac{J_{n}}{n}(\mu^{-n}-t^{n})
\right]{\cal G}^{(m,n)}|0\rangle}
{(1-\mu t)\langle0|{\cal G}^{(m,n)}|0\rangle}\right|_{\displaystyle{t=0}}.
\end{equation}
Since in our case $\langle0|{\cal G}|0\rangle= \epsilon$ and
\begin{equation}\label{11}
\frac{\langle0|\exp\left[\sum\frac{J_{n}}{n}(\mu^{-n}-t^{n})\right]
{\cal G}^{(m,n)}|0\rangle}
{\langle0|{\cal G}^{(m,n)}|0\rangle}=
1+\frac{1}{\epsilon}\mu^{-n-1}t^{m-1}(1-\mu t),
\end{equation}
it is obvious that
\begin{equation}\label{12}
\phi_{i}^{(m,n)} = \mu^{i-1}+ \frac{1}{\epsilon}\mu^{-n-1}\delta_{mi}.
\end{equation}
Therefore, the basic vectors are singular in the limit $\epsilon\rightarrow 0$,
but
from {\it normalized} $\tau $- function we can obtain that the limit of
the "Japanese"
correlator
\begin{equation}\label{13}
\langle0|\exp\left[\sum J_{n}T_{n}\right]{\cal G}^{(m,n)}|0\rangle=
\epsilon \frac{\det[\psi^{(m,n)}_{i}(\mu_{j})]}{\Delta(\mu)}
\end{equation}
has well-defined limit when $\epsilon \rightarrow 0$. For example, in
the case of two
Miwa variables ($N=2$) (it is trivial to check it for general case and
reproduce
proper expression)  from eq.(6)
\begin{eqnarray}
\langle0|\exp\left[\sum\frac{J_{n}}{n}
(\mu^{-n}_{1}-\mu^{-n}_{2})\right]{\cal G}^{(m,n)}|0\rangle = \nn\\
 = \epsilon + \frac{1}{\Delta(\mu)}
\left\{\delta_{m,1}\left|
\begin{array}{cc} \mu_{1}^{-n-1} & \mu_{2}^{-n-1}\\ \mu_{1} & \mu_{2}
\end{array}\right|- \delta_{m,2}\left|
\begin{array}{cc} \mu_{1}^{-n-1} & \mu_{2}^{-n-1}\\ 1 & 1
\end{array}\right|\right\}\label{14}
\end{eqnarray}
It coincides with expression (\ref{elchar}) and stress that it is impossible to
do the
regularization procedure under the determinant ($i.e.$ to regularize just the
basis
vectors giving the point of the Grassmannian).

Now let us stress that we determined the determinant form of the singular
$\tau$-function. The crucial difference with the regular case is the
possibility to "skip" some vectors $\phi_i(\mu) \sim \mu^{i-1} + O(1/\mu)$ in
the determinant (see the next subsection). The total lenght of the skipped
"gap" $S$ is an important parameter \cite{SW}. It determines, in particular,
the asymptotics of singular $\tau$-functions at small $t_1$ when all other
times are equal to zero:

\beq\label{SegWil}
\tau(t_1) \sim t_1^S.
\eeq

\subsection{Character expansion of generic $\tau$-function}

Now we would like to discuss more on the statement {\it iii)} to involve it
into the framework of generic KP $\tau$-functions. To start, let us consider
an arbitrary $\tau$-function in the determinant form (\ref{detrep}) with some
vectors

\beq\label{*}
\phi_i=\mu^i\sum_{k\ge 1} p_{ik}\mu^{-k}.
\eeq
Let us consider for a while the determinant (\ref{detrep}) of the finite size
$N\times N$. Then, after permuting the columns in the determinant and
multiplying this by $\mu^{-N}$, one can get

\beq\label{**}
\tau = \mu^{N} {\det \bar \phi_i (\mu_j)\over \Delta (\mu)},
\eeq
where

\beq\label{c1}
\bar \phi_i (\mu) \equiv \sum_{k=1} \bar p_{ik} \mu ^{-k},
\eeq
and
\beq\label{c2}
\bar p_{ij} \equiv p_{N-i+1,j-i+1},\ \ p_{ij} \equiv 0 \ \ \hbox{for}\ \  j<1.
\eeq
Then we use the formula

\beq
{\rm det}_{ij} \left( \sum_{k =1}^{\infty} f_{ik}p_{jk} \right)
= \sum_{1\leq k_1 < k_2 < ...}^{\infty} {\rm det}_{ij} f_{ik_j}
\ {\rm det}_{ij} p_{i k_j}\label{maineq}
\eeq
to rewrite $\tau$-function (\ref{**}) as a sum

\ba\label{15}
\tau = \sum_{1< k_1 < k_2 < ...}^{\infty} \det_{ij} \bar p_{i,k_j}
\ {\det_{ij}\mu_i^{N-k_j-1} \over \Delta (\mu)} = \\
= \sum_{m_1 \ge m_2 \ge ...\ge 0}^{\infty} \det_{ij} p_{i,m_j+i-j+1}
\ {\det_{ij}\mu_i^{-m_j+j-1} \over \Delta (\mu)},
\ea
and we replaced $k_i \longrightarrow m_{N-i+1}+N-i+1$ (compare with
(\ref{1}) and (\ref{5})--(\ref{6})) and reflected the second index in
the last determinant $i \longrightarrow N-i$ and analogously both indices
in the first determinant.

One can observe that expression (\ref{16}) is nothing but the linear
combination of the characters (like (\ref{elchar})) discussed in the
previous subsections. Certainly, it had to be evident from the very
beginning that any $\tau$-function can be expanded into a sum of the
characters as, in Miwa variables, any $\tau$-function is a symmetrical
function (to be a function of times) and the characters of groups
$GL(N)$ (= all possible Young tables) span the space of such functions.
Thus, the only question to be addressed to is what are the coefficients
$P\{m_i\}$'s of the expansion

\beq
\tau =
\sum_{m_1 \ge m_2 \ge ...\ge 0}^{\infty}  P\{m_i\}\chi_{\{m_i\}}(G).
\label{charexptau}
\eeq
In fact, one can just demand for the $\tau$-function to satisfy the
Hirota bilinear equations. Then, it can be demonstrated \cite{Hir} that
it means: the coefficients $P\{k_i\}$'s should be the Plucker coordinates
in the infinite dimensional Grassmannian. In its turn, this just implies
that they can be presented in the determinant form

\beq\label{Plucker}
P\{m_i\} = {\rm det}_{ij} p_{i,m_j+i-j+1}.
\eeq
Thus, we are led to formula (\ref{detrep}), there being no any
restrictions to the asymptotics of the vectors $\phi_i(\mu)$.

Now, one can see immediately from (\ref{Plucker}) that, if some
$m_{k+1}=0$ (and, therefore, so all $m_p,\ p>k$), then $P\{m_i\}$ will be
determinant of block matrix:

\beq\label{x1}
P\{m_i\} = \det \left|
\begin{array}{ccccccc}
&&&\vdots&&&\\
&{A}&&\vdots&&{0}&\\
&&&\vdots&&&\\
\cdots&\cdots&\cdots&\vdots&\cdots&\cdots&\cdots\\
&&&\vdots&1&&{0}\\
&{B}&&\vdots&&\ddots&\\
&&&\vdots&{\ast}&&1
\end{array}
\right| = \det_{k\times k} A =\left.\det_{ij} p_{i,m_j+i-j+1}\right|_{i,j\le
k}.
\eeq
Thus, we obtain that the final answer for the expansion (\ref{charexptau})
does not depend on $N$ (certainly not be this would implies impossibility
to rewrite the expansion (\ref{charexptau}) in terms of time variables), in
spite of it was impossible to avoid introducing $N$ for the calculations.
The only restriction to $N$ is that it should be larger than the number of
non-zero $m_i$'s, i.e. there should exist such $k\le N$ that the
representation with these $m_i$'s can be embedded into the group $GL(k)$.

For the future use, let us remark that
given some $P\{m_i\}$ one can always construct a whole family of
solutions:
\be
P_V\{m_i\} = e^{\sum_i V(m_i)} P\{m_i\}
\label{potinpluck}
\ee
with any function $V(x)$ (this corresponds to the change of $p_{ik}$
for $e^{V(k)}p_{ik}$).

To conclude the section, let us reformulate more precisely the statement
{\it iii)}. In fact, from (\ref{*})--(\ref{Plucker}) one can see that the
$\tau$-function is singular, i.e. there are gaps in its determinant
representation (\ref{detrep}) when at least one of the coefficients
$p_{i1}$ in (\ref{*})  is equal to zero. Actually, singular $\tau$-function
is implied to be zero at all times equal to zero, i.e. when all $\mu_i$ in
(\ref{15}) are infinite. At these values of $\mu_i$'s, the only trivial
character (when all $m_i=0$ in (\ref{15})) does contributes to the sum
(\ref{charexptau}). Therefore, $\tau$-function is singular, whenever the
trivial character does not contribute, i.e. (see (\ref{charexptau}) and
(\ref{Plucker})) $P\{m_i=0\} = \det_{ij}p_{i,i-j+1} = \prod_i p_{i1} = 0$
(note that the matrix $p_{i,i-j+1}$ is upper-triangle -- see (\ref{c2})).

Thus, we proved that

$iii')$ Arbitray $\tau$-function can be expanded to an (infinite) sum of
the characters of groups $GL(N)$ (\ref{4}) or (\ref{7}), the $\tau$-function
being singular if and only if at least one of $p_{i1}=0$.

This completes our consideration of the connection between (singular)
$\tau$-functions and characters of groups $GL(N)$.

\section{Bilinear combination of characters}

An obvious particular example of Plucker function $P_R = P\{k_i\}$ of the
form (\ref{Plucker}) is provided by characters themselves, $P_R \sim
\chi_R(\bar G^{-1})$ with some element $\bar G$.
Making use of the freedom (\ref{potinpluck}) we conclude that
\be
K(G,\bar G) \equiv
\sum_{1\leq k_1 < k_2 < ...}^{\infty} \chi_{\{k_i\}}(G) \chi_{\{k_i\}}
(\bar G^{-1})\ e^{\sum_i V(k_i)} =\\=
\sum_R \chi_R(G)\chi_R(\bar G^{-1})e^{V(R)}
\label{bilinchar}
\ee
is always a KP $\tau$-function as a function of time variables
$t_k = \frac{1}{k} {\rm tr} G^k={\f k}\sum_i \lambda^{k}_i=
{\f k}\sum_i \mu^{-k}_i$.\footnote{In fact, we have not chosen the
normalization of this $\tau$-function yet. We fix it to be
$\tau(0)=1$, which sometimes requires some trivial
renormalization of $K(G,\bar G)$.}

In order to construct $\tau$-function of Toda lattice hierarchy
one should introduce somehow the additional set of the independent parameters
(negative times). In formula (\ref{bilinchar}) these parameters are
$\bar t_{m} = \frac{1}{m} \hbox{tr}\bar G^{-k}={\f k}\sum_i \bar\lambda_i^{-k}=
{\f k}\sum_i \xi_i^k$.
We establish below that a particular
bilinear combinations of characters (\ref{bilinchar})
does really solve to Toda lattice hierarchy.

The simplest example of such solution which has the apparent
group-theoretical meaning is $e^{V(k)}={\f k!}$. Then
\beq\label{L1-22}
\tau (t,\bar t) = \frac{\det \exp [\lambda_{i} \xi_{j}]}
{\Delta(\lambda ) \Delta(\xi )}
\eeq
is nothing but Itzykson-Zuber formula (\ref{IZ}).

We can immediately
calculate the basic vectors using formulas of section 5.4.
After proper normalization these are
\beq\label{L1-47}
\phi_{i}(\mu ) = \mu^{i-1} \sum_{k=0}^{\infty}e^{V(k+N-1)-V(N-i)}
P_{k}(\bar t)\mu^{-k} \stackrel{\mu \gg 1}{\sim } \mu^{i-1}(1-O(\mu^{-1}).
\eeq

Let us discuss properties of expression (\ref{bilinchar})
as the $\tau$-function.
In particular, we will find an
explicit representation of the
corresponding Clifford element in the fermionic correlator (\ref{tau}).
One can consider this element as parametrized by the whole set of the
negative times; in this case correlator takes the form
\beq\label{L2-1}
\tau_n (t,\bar t) \equiv \langle n| e^{\displaystyle{H(t)}} \ {\cal G}(\bar t),
\ |n \rangle
\eeq
where
\beq\label{L2-2}
{\cal G}(\bar t) \equiv {\cal G}\times e^{\displaystyle{\bar H(\bar t)}}.
\eeq
Therefore, in Miwa variables, one can obtain the determinant representation
(\ref{detrep}) where the basic vectors are now the functions of $\{\bar t\}$:
\beq\label{L2-3}
\tau_n(t,\bar t) = \langle n|{\cal G}(\bar t)|n\rangle \
\frac{\det \phi_{i}(\mu_{j})}{\Delta(\mu )},
\eeq
where $\phi_{i}(\mu )$ have the following form:
\beq\label{L2-4}
\phi_{i}^{\it{(can)}}(\mu ) = \mu^{n-1} \ \frac{\langle n|\psi_{n-i}
\psi^{\ast}(\mu )\
{\cal G}(\bar t)|n\rangle}{\langle n|{\cal G}(\bar t)|n\rangle }.
\eeq
Here superscript "can" means that these vectors has the {\it canonical} form:
\beq\label{L2-5}
\phi_{i}^{\it{(can)}}(\mu ) = \mu^{i-1} + \sum_{k=0}^{\infty}\mu^{k-1}
\frac{\langle n|\psi_{n-i}\psi^{\ast}_{n+k}\ {\cal G}(\bar t)|n\rangle }
{\langle n|{\cal G}(\bar t)|n\rangle}.
\eeq

In order to reproduce (\ref{L1-47}) one can use a very simple (non-ordered !)
Clifford element which describes completely the point of Grassmannian which
corresponds to "group-theoretical" $\tau$-function (\ref{bilinchar}):
\beq\label{L2-7}
{\cal G} = \exp \{\sum_{k\in Z} V_{k}\psi_{k}\psi^{\ast}_{k}\}
\equiv \prod_{k\in Z} \ {\cal G}^{(k)},
\eeq
where
\beq\label{L2-8}
{\cal G}^{(k)} \equiv e^{\displaystyle{V_{k}\psi_{k}\psi^{\ast}_{k}}} =
1+(e^{\displaystyle{V_{k}}}-1)\psi_{k}\psi^{\ast}_{k}.
\eeq
Then it is easy to see that fermionic modes are transformed under the
action of this Clifford element {\it almost} trivially
\beq\label{L2-9}
{\cal G}\psi_{k}{\cal G}^{-1} = e^{\displaystyle{V_{k}}}\psi_{k} \ , \
{\cal G}\psi^{\ast}_{k}{\cal G}^{-1} = e^{-\displaystyle{V_{k}}}\psi^{\ast}_{k}
\eeq
and in this case the canonical basic vectors (\ref{L2-5}) have the form:
\beq\label{L2-10}
\phi_{i}^{\it{(can)}}(\mu ) = \sum_{k=0}^{\infty} \mu^{i+k-1}
e^{\displaystyle{[V_{n-i+k}-V_{n-i}]}}\sum_{m=0}^{i-1}
P_{m}(-\bar t)P_{k-m}(\bar t).
\eeq
One can show that combining columns in the
determinant it is possible to obtain the new
(non-canonical) basis
\beq\label{L2-11}
\phi_{i}(\mu ) = \sum_{k=0}^{\infty} \mu^{i+k-1}
e^{\displaystyle{[V_{k+n-i}-V_{n-i}]}} \ P_{k}(\bar t),
\eeq
since these vectors are linear combinations of the canonical ones:
\beq\label{L2-12}
\phi_{i}(\mu ) = \sum_{k=1}^{i} P_{i-k}(\bar t) \
e^{\displaystyle{[V_{n-k}-V_{n-i}]}} \ \phi_{k}^{\it{(can)}}(\mu ).
\eeq
The vectors (\ref{L2-11}) coincide with (\ref{L1-47}), the parameter $N$
playing the role of zero time $n$. It means that we really have demonstrated
that formula (\ref{bilinchar}) is the $\tau$-function of Toda
lattice hierarchy.

As the simplest example of the above construction,
for the trivial element ${\cal G}=1$ $\tau$-function
(\ref{L2-1}) has well-known form
\beq\label{L2-13}
\tau_n (t,\bar t) = \exp \{\sum_{k=1}^{\infty}kt_{k}\bar{t_{k}}\}
\eeq
and (non-canonical) basic vectors are:
\beq\label{L2-14}
\phi_{i}(\mu ) = \mu^{i-1}
\exp \{\sum_{k=1}^{\infty}\bar t_{k}\mu^{-k}\}.
\eeq

Strictly speaking, the element (\ref{L2-7}) has (for some particular
choises of
$V_{n}$) the {\it infinite} (or, vice versa, {\it zero}) norm since
\beq\label{L2-15}
{\cal G}|n\rangle = \prod_{k=-\infty}^{-1} e^{\displaystyle{V_{k}}}|n\rangle.
\eeq
This is of great importance, however, since the ratio $\tau(t,\bar t)/
\langle n|{\cal G}(\bar t)|n\rangle $ in (\ref{L2-3}) as well as the ratio in
(\ref{L2-5}) is {\it finite}. To avoid all the troubles, one can trivially
observe
that there exists {\it normally ordered} element (we use the first normal
ordering of section 5) with the same
transformation properties as in (\ref{L2-9}) which has {\it finite}
norm from the very beginning:
\beq\label{L2-16}
\tilde {\cal G} = \ddagger\exp \{ \sum_{k<0}(1-e^{\displaystyle{-V_{k}}})
\psi_{k}\psi^{\ast}_{k} + \sum_{k\geq 0}(e^{\displaystyle{V_{k}}}-1)
\psi_{k}\psi^{\ast}_{k}\}\ddagger
\eeq
and
\bea\label{L2-17}
\tilde {\cal G}|n\rangle = \frac{\prod_{k=-\infty}^{n}
e^{\displaystyle{V_{k}}}}
{\prod_{k=-\infty}^{-1} e^{\displaystyle{V_{k}}}}\ |n\rangle \equiv \nn \\
\equiv \left \{
\begin{array}{ll}
\prod_{k=0}^{n} e^{\displaystyle{V_{k}}} \ |n\rangle  & n\ge 0 \\
\prod_{k=n}^{-1} e^{\displaystyle{-V_{k}}} \ |n\rangle  & n< 0
\end{array} \right.
\eea
Two elements (\ref{L2-7}) and (\ref{L2-16}) are related by the obvious
expression:
\beq\label{L2-18}
{\cal G}= \tilde {\cal G} \ \prod_{k=-\infty}^{0}e^{\displaystyle{V_{k}}}.
\eeq

It is important to stress that (\ref{bilinchar}) describes KP hierarchy
w.r.t. time variables $t_k$ at any choices of the variables $\bar t_k$.
In particular, the case of one hole correlator in $2d$ YM theory (see
section 3) corresponds to all $t_k=1$. In this case,
the basis vectors (\ref{L2-11})
depend on the number of Miwa variables $N$. It would mean that such
correlator is not any KP $\tau$-function, despite of the construction above.
The puzzle is resolved by noting that the basis vectors depend only on the
number of variables $\xi_i$, which describe negative times, and
their number is equal to the number of $\lambda_i$ only by theoretical-group
reasons.

Let us note now that the function
$V(R) = \sum_i V(k_i)$ being represented in the form
\be\label{pott}
V(R) = \sum_{n=1}^{\infty}  s_n \left(\sum_i k_i^n\right)
\ee
is very similar to
\be
\tilde V(R)
 = \sum_{n=1}^{\infty}\tilde s_n C_n(R),
\ee
where $C_n(R)$ is the value of $n$-th Casimir operator in representation
$R$. However, the exact form of $\tilde V(R)$ depends on the choice of Casimir
operators. Canonical choice, related to the equation
\be
\Delta_n \chi_R = C^c_n(R)\chi_R,
\label{chareq}
\ee
is:
\be
C^c_n(R) = {\rm tr} _R E_{i_1 i_2}...E_{i_n i_1}
\ee
and does not provide $C_n(R)$ in the form, necessary for (\ref{bilinchar}).
Operators defined so that
\be
C^f_n(R) = \sum_i k_i^n
\ee
are also Casimir operators (they are non-linear functions of $C^c_n(R)$),
they are suitable for the use in (\ref{bilinchar}), but they do not arise in
eq.(\ref{chareq}). The special case is quadratic Casimir operator for $SL$
group (not $GL$): $C^f_2(R) - C^c_2(R) = const$.

Formula (\ref{bilinchar}) arises as partition function of $G(KM)^2$ with one
site $\alpha$ and two external fields $L$ and $\bar L$, $G = e^L$, $\bar G
= e^{\bar L}$, if it is considered as "unitary" model (i.e with {\it sums}
over integer eigenvalues of $\Phi$ instead of integrals).  It also arises
as correlator of two Wilson loops on a sphere in the $2d$ Yang-Mills
theory \cite{Mig,Wit}, in this case usually $V(R) = sC^c_2(R)$ (for
generic potential $V_R$ one can speak about "generalized $2d$ Yang-Mills
model").  We see that it is better to define it with potential $V(R)$
rather than $\tilde V(R)$ : then this quantity is KP $\tau$-function.
Moreover, it is automatically a Toda-lattice $\tau$-function, if "negative
times" are introduced as $\bar t_k = \frac{1}{k} {\rm tr}\bar G^{-k}$.

\section{Comments on Kontsevich model}

Eq.(\ref{bilinchar}) naturally arises from "unitary" (i.e. discretized)
Kontsevich model with single site $\alpha$ and two background fields. A
slightly more general model with $m$ background fields is
\beq
{\cal F}_V\{L\} \equiv {{\cal Z}_V\{L\} \over {\cal C}_V\{L\}} = \int
d\Phi \prod_{\mu =1}^m [dU_\mu] e^{{\rm tr}V(\Phi)}
\prod_{\mu =1}^m e^{{\rm tr} L_\mu U_\mu\Phi U_\mu^\dagger}=
$$
$$
= \prod_i \int d\phi_i e^{V(\phi_i)}\Delta(\phi)^{2-m} \prod_{\mu = 1}^m
\hat\chi_{\{\phi_i\}}(e^{L_\mu}),\label{16}
\eeq
where $\hat\chi_{\{\phi_i\}}(e^{L_\mu}) = \frac{\Delta(e^L)}{\Delta(L)}
\chi_{\{\phi_i\}}(e^{L_\mu})$.
The sewing operation, i.e. integration over $g$ pairs of background
fields with the weight $\delta(L_1+ L_2)dL_1 dL_2$  gives:

\be
{\cal F}_V\{L\} =
\prod_i \int d\phi_i e^{V(\phi_i)}
\Delta(\phi)^{2-2g-m} \prod_{\mu = 1}^m
\hat\chi_{\{\phi_i\}}(e^{L_\mu}),\label{17}
\ee

The discretized ("unitary") version of the model is

\ba
{\cal F}_V^{dscr}\{L\}  = \prod_i \sum_{\phi_i\in Z_+}  e^{V(\phi_i)}
\Delta(\phi)^{2-2g-m} \prod_{\mu = 1}^m
\hat\chi_{\{\phi_i\}}(e^{L_\mu}) =\\=
\sum_R e^{V(R)} d_R^{2-2g-m} \prod_i \hat\chi_R(e^{L_i}),
\label{Kondis3}
\ea
An important feature of the discrete sum is that it is defined over {\it
model} space and thus eigenvalues $\phi_i$ are not only restricted to be
integers, but also they should be {\it non-negative} integers. This makes
the discretized model essentially different from continuous Kontsevich
model.  Namely, in the case of continuous GKM one could usually rely upon
quasiclassical (steepest descent) calculation, which gives rise to
expansion of the integral in powers of $m_i^{-1}$ with matrix $M$ related
to $L$ as $V'(M)=L$.
However, this regime can be forbidden in discretized model, if the stable
point of the action lies in the domain of {\it negative} $\phi_i$ (it
depends on the potential). This
implies that besides the Kontsevich-type "phase" the discretized model
has also another - simpler - phase, where the natural expansion is just in
powers of $G_\mu = e^{L_\mu}$. We refer to it as to "character
phase". It is in this phase that our reasoning from the previous section
was actually applicable and this was why the point of Grassmannian
appeared so simple.

The study of Kontsevich phase is more complicated. Indeed, it corresponds
to modified reduction condition (having now non-local form) and the Ward
identity which usually fixes the string equation contains now some boundary
terms. But we would like to stress that in this phase the basis vectors for
the KP case (in contrast to the character phase) do not depend on $N$ from
the very beginning (the reason for this is that the point of the
Grassmannian given by ${\cal G}(\bar t)$ (\ref{L2-2}) does not depend on the
number of Miwa variables $\xi_i$ when all times $t_k=1$).

In character phase the quantities (\ref{Kondis3}) are $\tau$-functions
only for $g=0$ and $m=1,2$.
What happens in other cases can be illustrated by the following simple
example.  Take $m=2$ and $V = 0$, i.e. ${\cal G}=1$ (see (\ref{L2-13})).
Let us begin from the $1\times 1$ matrices. Then we always have (for any
$g$)
\be\label{18}
d\lambda^{{\f 2}}d\bar \lambda^{-{\f 2}}K =
d\lambda^{{\f 2}}d\bar \lambda^{-{\f 2}}\sum_{k=0}^{\infty}
\lambda^k\bar\lambda^{-k} = \frac{d\lambda^{{\f 2}}d\bar \lambda^{{\f 2}}}
{\lambda -\bar \lambda} = \langle 0 \mid \psi(\lambda)
\psi^{\ast}(\bar\lambda) \mid 0 \rangle =\Psi(\lambda,\bar\lambda).
\ee
Now we have a simple prescription (\ref{Toda}) to define the quantity for
$N\times N$
matrices so that it automatically becomes $\tau$-function (certainly, this
is equivalent to formula (\ref{L2-13}) when is rewritten in Miwa
variable):
\be\label{19}
K_N = \frac{{\rm det}_{ij}K_1(\lambda_i, \bar\lambda_j)}{\Delta(\lambda)
\Delta(\bar\lambda)}={\prod_i \bar\lambda_i\over \prod_{i<j}(\lambda_i -
\bar\lambda_j)}={\Psi(\lambda_i,\bar\lambda_j)\over\Delta(\lambda)\Delta
(\bar\lambda)},
\ee
where we used the formula for the Cauchy determinant:
\be
\det_{ij}{\f x_i-y_j} = {\Delta(x)\Delta(y)\over\prod_{i,j}(x_i-y_j)}.
\ee
One can easily see that this leads to the formula with $g=0$ (and
non-vanishing potential can be easily introduced, as explained in the
previous section).

This is already enough to conclude that expression for
$g > 0$ are not $\tau$-functions, simply because they are different from
the one arising by the above rule.  To see what they look like let us
consider an example of $N=2$ (still with  $V=0$ and $m=2$).
Then we have:
\be\label{20}
K_2=\sum_{k_2>k_1\ge 0}^{\infty}{(\lambda_1^{k_1}\lambda_2^{k_2}-
\lambda_1^{k_2}\lambda_2^{k_1})(\bar\lambda_1^{k_1}\bar\lambda_2^{k_2}-
\bar\lambda_1^{k_2}\bar\lambda_2^{k_1})\over (k_1-k_2)^{2g}}=\\
= {\bar\lambda_1\bar\lambda_2\over\lambda_1\lambda_2-
\bar\lambda_1\bar\lambda_2}\left[Li_{2g}\left({\lambda_1
\over\bar\lambda_1}\right)+Li_{2g}\left(
{\lambda_2\over\bar\lambda_2}\right)-
Li_{2g}\left({\lambda_1\over\bar\lambda_2}\right)-
Li_{2g}\left({\lambda_2\over\bar\lambda_1}\right)
\right],
\ee
where $\displaystyle{Li_{p}[x]\equiv\sum_{i=1}^{\infty}{x^i\over i^p}}$
is nothing but polylogarithm function. It can be also
re-written as
\be
K_2=\int^{1}{dz_{2g}\over z_{2g}}\ldots\int^{z_3} {dz_1\over z_2}
\int^{z_2}{dz_1\over z_1}
\det\left|
\begin{array}{cc}
\Psi(\lambda_1,z_1\bar\lambda_1)&\Psi(\lambda_1,{\f z_1}\bar\lambda_2)\\
\Psi(\lambda_2,z_1\bar\lambda_1)&\Psi(\lambda_2,{\f z_1}\bar\lambda_2)
\end{array}\right|,
\ee
i.e. for the case of generic $N$
\be\label{21}
K_N=\left.J \frac{{\rm det}
\Psi(\lambda_i, z_j\bar\lambda_j)}{\Delta(\lambda)\Delta(\bar\lambda)}
\right|_{z_{ij}=1}.
\ee
Here $z_j = \prod_{k\neq j} z_{jk}$, $z_{kj} = z_{jk}^{-1}$ and $J$ is an
integral operation $J = \prod_{i<j} J_{z_{ij}}^{2g}$, $J_z =
\int\frac{dz}{z}$,
where integral is defined so that $J_z z^p = \frac{z^p}{p}$ for any $p$.

In other words, for $g\neq 0$ in the character phase we get integrals of
$\tau$-functions. Indeed, it is not so suprising and one can trivially
understand this already for the case of multi-point correlators in the
sphere (see (\ref{Kondis3})). It is sufficient to use the identity for
characters:

\beq\label{22}
\chi_\alpha (G_1)\chi_\alpha (G_2) = d_
\alpha \int [dU] \chi_\alpha[G_1UG_2U]
\eeq
to reduce the arbitrary
$n$-point correlator to $(n-2)$-integrated $\tau$-function of 2TDL
hierarchy. In this case, similar to eq.(\ref{21}),
the integrations go over the variables
which "rotate" the group element $G$ giving Miwa times.
We do not know what does it mean in the integrability
framework.

\section{Concluding remarks}

The simple manipulations, described in the previous subsection, raise many
puzzling problems, which can stimulate new investigation and after all
provide a new view on the entire theory of matrix models and
integrability. Now we want to mention some of these problems somewhat more
explicitly.

{\it Nature of variables in the $G(KM)^2$}

Once external (background)
fields are introduced as in (\ref{g2km}), one is naturally led to
interpretation of integration over eigenvalues of $\Phi^{(\alpha)}$-fields
as that over weight space of $GL$ group. This weight space can be also
called the space of all represntations, i.e. a {\it model} space. This
gives rise to a natural suggestion to consider a "unitary-group" version
of the {\it model}, where integration is substituted by a sum over
irreducible representations of unitary groups, i.e. over Young diagramms.
Then $G(KM)^2$ includes integration over {\it model}-valued fields, and what
is lacking is the group-theoretical meaning (if any) of the model
(\ref{g2km}), which implies a somewhat specific interaction between
different {\it model}-variables $\Phi^{(\alpha)}$. This problem is of
course absent in the case of a single point $\alpha$, i.e. for Kontsevich
model with several external fields, which can be in fact sufficient for
the study of the generalized $2d$ Yang-Mills and sum other theories.

Alternatively one could think that $e^{\Phi^{(\alpha)}}$ should be
interpreted as elements of unitary group rather then label
representations. This is a different view, since it implies that
eigenvalues of $\Phi^{(\alpha)}$ lie on a segment $[0,2\pi]$ which is
treated as compactified circle. One of the problems with this
interpretation  is that it seems to require some periodicity under the
shift of any eigenvalue of $\Phi^{(\alpha)}$ by $2\pi$,  which is
apparently absent in Lagrangain of the model (\ref{g2km}). Eq.(\ref{IZ})
could serve as an illustration that this periodicity (for eigenvalues of
$L$) can be restored after integration over $U$ (provided eigenvalues of
$X$ are integers, i.e. for representations of {\it unitary} group),
however eq.(\ref{gkm}) seems to imply that this does not need to be the
case in the $G(KM)^2$, as defined in eq.(\ref{g2km}). An even more severe
problem is apparent non-priodicity of potential term ${\rm tr}V(\Phi)$ for
polinomial $V$.\footnote{
There is some discussion of this issue in a recent paper \cite{CAMP} in
the specific case of quadratic potential.}
Because of these problems we prefer the above interpretation of
$\Phi^{(\alpha)}$'s as elements of the weight ({\it model}) space, when
$V(\Phi)$ has a natural interpretation  in terms of Casimir operators. The
alternative picture with unitary $e^\Phi$ can arise after certain "Fourier
transform" of the model (\ref{g2km}), which remains to be worked out (it
has very good chances to be similar to the same model (\ref{g2km}) in the
Gaussian case of quadratic potential $V(\Phi) \sim \Phi^2$.)

{\it Integrability properties}

Conceptually integrability is usually
related to correlation functions of (any kind of) free fields (i.e.
allowing for Wick theorem) on (any kind of) Riemann surfaces: such
correlators can be always interpreted as $\tau$-functions in Miwa
parametrization (usually one considers only free-fermion $\tau$-function,
related to level $k=1$ Kac-Moody algebras, but this should not be a
serious restriction\footnote{
Such restriction is an explicit consequence of requirement that integrable
flows {\it commute} instead of forming some closed non-abelian algebra.}).
However, amplitudes in topological and string models are usually {\it
integrals} of such correlation functions over some moduli (of Riemann
surfaces or bundles over Riemann surfaces), essentially over
whole Universal Moduli Space, but not over its subspaces corresponding to
finite genus Riemann surfaces. Thus, generic quantities of
interest in this framework are integrals of $\tau$-functions (unless in
some special circumstances like (\ref{bilinchar}) and (\ref{19}) where
integrals
accidently are not present). Adequate group-theory (or orbit-theory)
interpretation of integrals of $\tau$-functions (similar to that in
\cite{Kac} for $\tau$-functions themselves) remains to be found.

This is a trivial remark, what deserves emphasising is that from certain
point of view it is a step back from the main achievement of matrix model
theory. Namely, matrix models were used to prove that exact partition
functions, if considered as functions of coupling constants (i.e.
as functions on the spaace of string models or of moduli of bundles over
Riemann surfaces) are $\tau$-functions, i.e. these were $\tau$-functions
arising as correlators in spectral space, not on the world surfaces, and
thus the above argument did not necessarily implied ocuurence of integrals
of $\tau$-functions. Remarkably, in the framework of Kontsevich models
dependence on the string model could be imitated in effective way by that
on the matrix external field $L$.  Somehow, we see that this is no longer
true for the $G(KM)^2$: one should abandon at least one of the two
suggestions: (a) that $G(KM)^2$ is a natural generalization of GKM for
description of more sophisticated string models and/or (b) that
non-perturbative partition functions are usually $\tau$-functions. Option
(a) is for sure true, at least, in the sense that $G(KM)^2$ is still not
general enough. Most
probably both suggestions deserve serious modifications, but the study of
$G(KM)^2$ can shed more light on this problem.

{\it Specifics of Kontsevich model}

There is certain difference between the ways in which Miwa variables arise
in GKM and in our considerations of the Kontsevich phase in this paper. In
group (character) theory
the time-variables are usually represented as $t_k = \frac{1}{k}{\rm tr}_F
G^k = \frac{1}{k} {\rm tr} e^{kL}$, while in GKM $t_k = \frac{1}{k} {\rm
tr} M^{-k}$, where $M$ is related to $L$ in potential-dependent way: $V'(M)
= L$. This difference is because in the context of GKM the time-variables
appear in quasiclassical expansion of partition function around a stable
point $\Phi = L$ in the integral (\ref{gkm}), while in character theory
the idea is that every particular character is by itself a polinomial in
$t$-variables. These two approaches are in a sense orthogonal. The
correspondence can be restored by a change of reasoning in character
theory, as described in the section 2 above.

Another possibility is to
consider specific potential $V_0(\Phi) = -\Phi\log\Phi$, when $L = V_0'(M)
= -\log M$, and the two descriptions literally coincide.  This is a
remarkable version of GKM for many reasons, of which we mention only two.
The simplest heuristic remark is that it should describe the $(p,1)$ model
with $p = +0$, i.e. with $c = 1 - \frac{6(p-q)^2}{pq} = -\infty$, which
has good chances to be related to some simple physical theories, including
$2d$ Yang-Mills. This argument seems not least convincing than the one
distinguishing GKM with potential $V_{-1}(\Phi) = \log \Phi$ ($p = -1$) as
the one describing (a part of) the $c=25$ (i.e. essentially $c=1$) string
model, which is partly confirmed by explicit investigation in
\cite{Mar,DiMoPle,KM}. Both potentials $V_0$ and
$V_{-1}$ are beyond the region of $p>1$ where quasiclassical treatement
of GKM is valid, and there are certain corrections to standard GKM
theory. The second remark is that these potentials are very similar to two
remarkable choices: $V_\Gamma(\Phi) = -\log \Gamma(\Phi)$ and
$V_\psi(\Phi) = \log \psi(\Phi)$ respectively (here $\Gamma(x)$ is Euler
Gamma-function and $\psi(x) = \frac{\partial}{\partial x}\log\Gamma(x)$.
These are distinguished because
$e^{V_\Gamma(x)} = \frac{1}{\Gamma(x)}$ and $e^{V_\psi(x)} = \psi(x) =
\sum_{k=0}^{\infty} \frac{1}{x+k}$, and for special choices of integration
contours (around the singularities at negative $x$) the integrals over
$\phi$'s reduce to {\it sums} over discrete negative integer values of
$\phi$ - this is exactly what necessary for going from non-compact
version of (\ref{g2km}) to the compact one. It deserves emphasizing
that arising sums are not over {\it all} integers, but only over
integers of one sign, as required for enumeration of Young tables (with
non-negative rows lenghts) and irreducible representation of unitary
groups.

{\it Generalization of characters}

Quantum mechanics on group manifolds and coadjoint orbits is not exhausted
by formulas (\ref{IZ}) and (\ref{qmech}).
There exist various matrix-like integral representations of
matrix elements, more general than characters, namely, --  for zonal
spherical functions $\Phi_{\{k_i\}}(q)$.
In particular, there are intriguing Harish-Chandra
formulas like \cite{HC,OP} (we write down these for the zonal spherical
functions on the symmetric spaces with the root system $A_{n-1}$)
\be\label{23}
\Phi_{\{k_i\}}(q)=\int_u du \prod_{l=1}^{n-1}\Delta_l^{i{K_{n-l+1}-K_{n-l}
\over 2a}-ma}(ue^{2aq}u^{-1}),
\ee
where $q=\hbox{diag}(q_1,\ldots,q_n)$, $K=(K_1,\ldots,K_n)$, $\Delta_l$ are
the lowest corner minors of the matrix $ue^{2aq}u^{-1}$, $u$ is a
stationary subgroup of the corresponding symmetric space, $du$ is
invariant measure (normalized to 1) and $m$ is a number depending on the
type of the symmetric space.

Moreover, it turns out, in certain cases zonal spherical functions
possess determinantal
representations, which allow to interpret them as $\tau$-functions,
despite they are defined as matrix elements, not as partition functions.
The simplest examples\footnote{See \cite{OP} for more sophisticated
examples.} are zonal spherical functions on the symmetric spaces of type $A\
III \ = SU(r,t)/S(U(r)\oplus U(t))$. They are labeled by a set of integers
$\{k_i\}$ and given in the explicit {\it detrminant} form \cite{BK} (for
the sake of brevity we write down the result only for the compact spaces):

\beq\label{24}
\Phi_{\{k_i\}}(q) \sim {\det_{ij} F_{k_i}(\lambda_j) \over \Delta (\lambda)
\Delta (k^2)},
\eeq
where $\lambda \equiv \cos 2q$ in notations of \cite{OP}, $F_{k_i}(\lambda) =
P^{\alpha,\beta}_{\{m_i\}}(\cos 2q)$ are Jacobi polynomials and $\alpha=\mu_1+
\mu_2+{1 \over 2}$, $\beta=\mu_2- {1\over 2}$,
$m_i={1\over 2}k_i-{1\over 2}(\mu_1+2\mu_2)$. The constants $\mu_1$ and
$\mu_2$ can be taken arbitrary though the group values are $\mu_1=2(r-s);
\ r,s \in Z,\ \mu_2=2$.
We do not dwell upon
this interesting subject here, because its implications for the general
theory remains obscure for us. At least it can be considered as an
interesting {\it application} of DH-like theory and confirm the
suggestion, that everything, arising from DH-theorem should have
interpretation in terms of $\tau$-functions. We refer to \cite{NiemiTi}
and
\cite{NeGo} for some more details about this application (in \cite{NeGo}
also a natural interpretation in terms of Calogero model is discussed).

{\it Note on the formula $\sum_R e^{\sum s_k C_k(R)} \chi_R(G)\chi_R(\bar
G^{-1})$}.

Let us mention one more similarity of this formula (\ref{bilinchar}) to the
standard GKM model.

Partition function of GKM is a $\tau$-function subjected to a
set of additional equations. These equations form
a set of Virasoro (or $W$-) constraints and their explicit appearance
depends on the shape of potential function. Moreover, the derivatives w.r.t.
the coefficients of the potential are expressed in a single way through
time derivatives \cite{LGGKM}.

Analogous procedure can be
performed in the model considered in this paper.
Namely, one can differentiate (\ref{bilinchar}) w.r.t. parameters $s_n$
of potential function (\ref{pott}). This gives rise to the Casimir
eigenvalues $C_n$ in the sum
(\ref{bilinchar}):
\be\label{L}
{\partial\over\partial s_n} \sum_R e^{\sum t_k C_k(R)}
\chi_R(G)\chi_R(\bar G^{-1})=
\sum_R C_n(R)e^{\sum t_k C_k(R)}
\chi_R(G)\chi_R(\bar G^{-1}).
\ee
Alternatively, one can use (\ref{chareq}) to represent the r.h.s. as
result of the action of certain differential operators on the characters.
Explicit expressions for
these operators can be easily found in terms of
$\lambda_i$-derivatives.
After Miwa transformation, these
can be finally rewritten as differential operators in times $t_k$ (in the
spirit of \cite{MMM}).

Certainly, to generate some analogs of Virasoro (or $W$) constraints, one
also needs Ward identities. They should connect Casimirs and powers
of Miwa variables and are more complicated than the usual ones \cite{KMMMZ}
due to appearance of boundary terms (as the sum in (\ref{bilinchar}) runs
only over positive integers).

Having a set of constraints of Virasoro type, one can naturally
compare the parameters
$s_n$ with "potential times" of \cite{LGGKM} as the r.h.s. of the
eq.(\ref{L})
can be usually obtained by taking derivatives in times $t_n$. Along with
the constraint algebra it fixes the
dependence of (\ref{bilinchar}) both on times $t_n$ and on times $s_n$.

\section*{Acknowledgements}

We benefited from the discussions with E.Antonov, A.Gerasimov, A.Gorsky,
A.Losev, N.Nekrasov, A.Niemi, M.Olshanetsky, I.Polyubin, A.Zabrodin.

The main part of this report was written during our stay at Institute of
Theoretical Physics, Uppsala University, and we highly appreciate the
hospitality and support. The work of S.K., A.Mar. and A.Mir. is
partially supported by grant 93-02-14365 of the Russian
Foundation of Fundamental Research.

\section*{Appendix A. Fermionic language for $\tau$-functions}
\def\theequation{A\arabic{equation}}
\setcounter{equation}{0}

\subsection*{A1. Free fermions}

In this Appendix some notations and definitons on the fermionic representation
of $\tau$-functions are collected. The main references are
\cite{KMMMZ,DJKM,KMMM}.

We start with the anticommutation relations for the fermionic modes:

\begin{equation}\label{A1}
\{\psi_{i},\psi^{\ast}_{j} \} = \delta_{ij} \;\;,\hspace{0.3in} \{\psi_{i},
\psi_{j} \}
=\{\psi^{\ast}_{i},\psi^{\ast}_{j} \} = 0
\end{equation}

Totally empty vacuum $|+\infty \rangle$ is defined  by relations
\begin{equation}
\psi_{i}\;|+\infty \rangle = 0 \;\; , \;\; i \in Z
\label{eq:a}
\end{equation}
Then the "$n$-th" vacuum is defined as follows:
\begin{equation}\label{A2}
|n \rangle = \psi^{\ast}_{n} \psi^{\ast}_{n+1} \ldots |+\infty\rangle
\end{equation}
and satisfies the following conditions
\begin{equation}\label{A3}
\psi^{\ast}_{k}|n \rangle = 0 \;\; k\geq n \;\;\;, \;\;\; \psi_{k}|n
\rangle = 0 \;\; k<n
\end{equation}
Free fermionic fields
\begin{equation}
\psi(z) \equiv \sum_{i \in Z} \psi_{i}z^{i}\;\;\;,\;\;\;\psi^{\ast}(z)
\equiv \sum_{i \in Z}
\psi^{\ast}_{i}z^{-i}
\end{equation}
can be bosonized with the help of the free {\it bosonic} field $\phi(z)$:

\begin{eqnarray}
\phi(z) = x -  i p \log z +  i \sum_{k \in Z} \frac{J_{k}}{k}z^{-k} \\
\label{eq:phi}
\mbox{[{\it x,p}]= {\it i} }\;\; ; \;\; [J_{m},J_{n}]=m\delta_{m+n,0} \nonumber
\end{eqnarray}
as follows\footnote{In this Appendix the normal ordering is taken with respect
to the bosonic vacuum.}
\begin{eqnarray}
\psi(z)= :e^{{\displaystyle \ i \phi(z)}}:\  \equiv \hspace{1.0in} \nonumber \\
\equiv e^{{\displaystyle \ i x}}\;e^{{\displaystyle p\log z}}\;
\exp [\sum_{k=1}^{\infty}\frac{\bar J_{k}}{k}z^{k}] \times \exp
[-\sum_{k=1}^{\infty}\frac{J_{k}}{k}z^{-k}]
\label{eq:q}
\end{eqnarray}

\begin{eqnarray}
\psi^{\ast}(z)= z:e^{{\displaystyle - i \phi(z)}}:\  \equiv \hspace{1.0in}
\nonumber \\
\equiv ze^{{\displaystyle -\ i x}}\;e^{{\displaystyle -p\log z}}\;
\exp [-\sum_{i=1}^{\infty}\frac{\bar J_{k}}{k}z^{k}] \times \exp
[\sum_{i=1}^{\infty}\frac{J_{k}}{k}z^{-k}]\label{eq:q'}
\end{eqnarray}
where
\begin{equation}\label{A5}
J_{-k} \equiv \bar J_{k} \;\;\;, \;\;\; k>0
\end{equation}
Using definition eq.(\ref{eq:phi}) one can easily show that
\begin{equation}\label{A6}
:e^{{\displaystyle i \alpha \phi(z)}}::e^{{\displaystyle i \beta\phi(w)}}: =
(z-w)^{\alpha \beta}:e^{{\displaystyle i \alpha \phi(z)+ i \beta \phi(w)}}:
\end{equation}
and therefore
\begin{equation}\label{A7}
\psi(z) \psi^{\ast}(w) = \frac{w}{z-w}:e^{{\displaystyle i \phi(z)-i \phi(w)}}:
\end{equation}
Let us introduce the "Hamiltonians"

\begin{equation}\label{Ham1}
H(T) \equiv \sum_{k=1}^{\infty} T_{k}J_{k}
\end{equation}

\begin{equation}\label{Ham2}
\bar H(\bar T) \equiv \sum_{k=1}^{\infty} \bar T_{k}\bar J_{k}
\end{equation}

Then one can show \cite{DJKM} that the following \underline {most important
formulas} hold:
\begin{eqnarray}
\langle n|\psi(z)=z^{n-1}\langle n-1|\exp [H(T- \epsilon (z^{-1}))] \nonumber
\\
\equiv z^{n-1}\hat X(z,T)\ \langle n-1|e^{{\displaystyle H(T)}}\label{main1}
\end{eqnarray}

\begin{eqnarray}
\langle n|\psi^{\ast}(z)=z^{-n}\langle n+1|\exp [H(T+ \epsilon (z^{-1}))]
\nonumber \\
\equiv z^{-n}\hat X^{\ast}(z,T)\langle n+1|e^{{\displaystyle
H(T)}}\label{main2}
\end{eqnarray}
where
\begin{equation}\label{A8}
\hat X(z,T) = e^{\xi(z,T)} \; e^{-\xi(z,\tilde \partial_{T})}
\end{equation}
\begin{equation}\label{A9}
\hat X^{\ast}(z,T) = e^{-\xi(z,T)} \; e^{\xi(z,\tilde \partial_{T})}
\end{equation}

\subsection*{A2. The derivation of the formula (37)}

In this Appendix as an example we demonstrate the derivation of formula
(\ref{miwataunotilde}) \cite{KMU}.

Since
\begin{eqnarray}
\langle n-N| \psi^{\ast}(z_{1}) \ldots \psi^{\ast}(z_{N}) = \hspace{1.0in}
\nonumber \\
= (-1)^{N(N-1)/2} \prod_{i=1}^{N}z_{i}^{1-n} \hspace{0.03in} \Delta(z)
\langle n|\exp [H(\sum_{i=1}^{N}\epsilon(z_{i}^{-1}))]\label{A10}
\end{eqnarray}

and

\begin{eqnarray}
\psi(w_{N}) \ldots \psi(w_{1})|n-N\rangle = \hspace{0.4in} \nonumber \\
=  \prod_{i=1}^{N}w_{i}^{n-N} \hspace{0.03in} \Delta(w) \exp [H(\sum_{i=1}^{N}
\epsilon(w_{i}))] |n\rangle\label{A11}
\end{eqnarray}

then, obviously,

\begin{eqnarray}
\langle n-N|\psi^{\ast}(z_{1}) \ldots \psi^{\ast}(z_{N}){\cal G} \psi(w_{N})
\ldots
\psi(w_{1})|n-N\rangle = \nonumber \\
= (-1)^{N(N-1)/2} \Delta(z) \Delta(w) \hspace{0.03in}\prod_{i=1}^{N}
z_{i}^{1-n}w_{i}^{N-n}
\times \nonumber \\
\times \langle n|\exp [H(\sum_{i=1}^{N}\epsilon(z_{i}^{-1}))]{\cal G}\exp
[H(\sum_{i=1}^{N}
\epsilon(w_{i}))] |n\rangle\label{A12}
\end{eqnarray}
and, therefore, in the case of $N$ Miwa variables

\beq
\new
\begin{array}{c}
\tau(T,\bar T) =
 \frac{\langle n-N|{\cal G}|n-N\rangle}{\Delta (z^{-1})\Delta (w)} \times\nn\\
\times\det \frac{\langle n-N+1|e^{\sum \frac{1}{k}J_{k}z^{-k}} \;
{\cal G} \; e^{\sum \frac{1}{k}\bar
J_{k}w^{k}}|n-N+1\rangle}{\langle n-N|{\cal G}|n-N \rangle}\label{A13}
\end{array}
\eeq

\subsection*{A3. Determinant representations for $\tau$-functions}

In this Appendix we derive the determinant representation of $\tau$-functions
in terms of ordinary (not Miwa) times. Some relations used in the derivation
are necessary to obtain formulas of the section 5.2. We follow the line of
the paper
\cite{KMMM}.

Let  ${\cal G}$  be an arbitrary element of the Clifford group which does
not mixes
the $\psi$- and $\psi ^\ast $- modes :

\beq\label{A14}
{\cal G} = \ :\exp [\sum     {\cal A}_{km}\psi _k\psi ^\ast _m]: ,
\eeq
where  :  :  denotes throughout this Appendix the normal ordering with respect
to the Dirac vacuum
$|0\rangle $. Then it is well known \cite{DJKM,10} that

\beq\label{A15}
\tau _n(x,y) = \langle n|e^{H(x)}{\cal G}e^{-\bar H(y)}|n\rangle
\eeq
solves the two-dimensional Toda lattice hierarchy, $i.e$. is the solution to
the whole set of the Hirota bilinear equations. Any particular solution depends
only on the choice of the element  ${\cal G}$ (or, equivalently it can be
uniquely
described by the matrix  ${\cal A}_{km}$). From eqs.(\ref{A1}) one can conclude
that
any element in the form (\ref{A15}) rotates the fermionic modes as follows

\beq\label{A16}
{\cal G}\psi _k{\cal G}^{-1} = \psi _jR_{jk}\hbox{ , }  {\cal G}\psi ^\ast _k
{\cal G}^{-1} =
\psi ^\ast _jR^{-1}_{kj} ,
\eeq
where the matrix  $R_{jk} $ can be expressed through  ${\cal A}_{jk}$ (see
\cite{26}). We
will see below that the general solution (\ref{A15}) can be expressed in the
determinant form with explicit dependence of  $R_{jk}$ . In order to calculate
$\tau $-function  we need some more notations. Using commutation relations
(\ref{A1}) one can obtain the evolution of  $\psi (z)$ and  $\psi ^\ast (z)$ in
times $\{x_k\}$, $\{y_k\}$ in the form

\beq\label{A17}
\psi (z,x) \equiv  e^{H(x)}\psi (z)e^{-H(x)} = e^{\xi (x,z)}\psi (z)\hbox{  ,}
\eeq

\beq\label{A18}
\psi ^\ast (z,x) \equiv  e^{H(x)}\psi ^\ast (z)e^{-H(x)} =
e^{-\xi (x,z)}\psi ^\ast (z)\hbox{  ;}
\eeq

\beq\label{A19}
\psi (z,\bar y) \equiv  e^{\bar H(y)}\psi (z)e^{-\bar H(y)} =
e^{\xi (y,z^{-1})}\psi (z)\hbox{ , }
\eeq

\beq\label{A20}
\psi ^\ast (z,\bar y) \equiv  e^{\bar H(y)}\psi ^\ast (z)e^{-\bar H(y)} =
e^{-\xi (y,z^{-1})}\psi (z)\hbox{ ,}
\eeq
where

\beq\label{A21}
\xi (x,z) = \sum ^\infty _{k=1}x_kz^k\hbox{ .}
\eeq
Using the definition of the Shur polynomials $P_k(x)$ (\ref{Shur}),
from eqs. (\ref{A17})-(\ref{A20}) one can easily obtain the evolution of the
fermionic modes:

\beq\label{A22}
\psi _k(x) \equiv e^{H(x)}\psi _ke^{-H(x)}
=\sum ^{\infty}_{m=0} \psi _{k-m}P_m(x) ,
\eeq

\beq\label{A23}
\psi ^{\ast }_k(x) \equiv e^{H(x)}\psi _k^{\ast }e^{H(x)} =
\sum ^{\infty}_{m=0} \psi ^{\ast }_{k+m}P_m(-x) ;
\eeq

\beq\label{A24}
\psi _k(\bar y) \equiv e^{\bar H(y)}\psi
_ke^{-\bar H(y)} =\sum ^{\infty} _{m=0} \psi
_{k+m}P_m(y);
\eeq

\beq\label{A25}
\psi _k^{\ast }(\bar y) \equiv e^{\bar H(y)}\psi
_k^{\ast }e^{-\bar H(y)}=\sum ^{\infty} _{m=0} \psi
_{k-m}^{\ast } P_m(-y).
\eeq
It is useful to introduce the totally occupied state  $|-\infty \rangle $
which satisfies the
requirements (c.f. with (\ref{eq:q}))

\beq\label{A26}
\psi ^\ast _i|-\infty \rangle  = 0\hbox{  , }  i \in  {\bf Z}\hbox{  .}
\eeq
Then any shifted vacuum can be generated from this state as follows:

\beq\label{A27}
|n\rangle  = \psi _{n-1}\psi _{n-2} ...|-\infty \rangle \hbox{  .}
\eeq
Note that the action of an any element  ${\cal G}$  of the Clifford group
(and, as
consequence, the action of $e^{-\bar H(y)} $) on  $|-\infty \rangle $  is very
simple: ${\cal G}|-\infty \rangle  \sim  |-\infty \rangle $, so using
(\ref{A23})
and
(\ref{A24}) one can obtain from eq.(\ref{A15}):

\beq
\new
\begin{array}{c}
\tau _n(x,y) =
\langle -\infty |...\psi ^\ast _{n-2}(-x)\psi ^\ast _{n-1}(-x){\cal G}
\psi _{n-1}(- \bar y)\psi _{n-2}(-\bar y) ...|-\infty \rangle  \sim \\
\sim \det [\langle -\infty |\psi ^\ast _i(-x){\cal G}
\psi _j(-\bar y){\cal G}^{-1}|-\infty
\rangle ]\left| _{i,j \leq  n-1}\right.\hbox{ .}
\end{array}\label{A28}
\eeq
Using (\ref{A16}) it is easy to see that

\beq\label{A29}
{\cal G}\psi _j(-\bar y){\cal G}^{-1} =
\sum _{m,k} P_m(-y)\psi _kR_{k,j+m}
\eeq
and the ``explicit" solution of the two-dimensional Toda lattice has the
determinant representation:

\beq\label{A30}
\tau _n(x,y) \sim \left. \det \ {\hat H}_{i+n,j+n} (x,y)
\right| _{i,j<0}\hbox{ ,}
\eeq
where

\beq\label{A31}
{\hat H}_{ij}(x,y) = \sum _{k,m} R_{km}P_{k-i}(x)P_{m-j}(-y)\hbox{  .}
\eeq
The ordinary solutions to KP hierarchy \cite{DJKM} correspond to the case
when the
whole evolution depends only of positive times $\{x_k\}$; negative times
$\{y_k\}$ serve as parameters which parameterize the family of points in
Grassmannian and can be absorbed by re-definition of the matrix $R_{km}$. Then
$\tau $-function of (modified) KP hierarchy has the form

\beq\label{A32}
\tau _n(x) = \langle n|e^{H(x)}{\cal G}(y)|n\rangle \sim \det [\sum  _k
R_{k,j+n}(y)P_{k-i-n}(x)]\left| _{i,j<0}\right.\hbox{ ,}
\eeq
where  ${\cal G}(y) \equiv  {\cal G}e^{-\bar H(y)}$ and

\beq\label{A33}
R_{kj}(y) \equiv  \sum  _m R_{km}P_{m-j}(-y)\hbox{  .}
\eeq

\section*{Appendix B. Higher Casimirs}
\def\theequation{B\arabic{equation}}
\setcounter{equation}{0}

In this Appendix we discuss some formulas on higher Casimirs which appear in
the formulas of the section 6.

We start with the definitions from the book \cite{Zhelobenko}.
For GL(N) the standard Casimirs $C_{n}$ have the form Eq.(50). Let, as before,
$\alpha \equiv (m_{1}, \ldots , m_{N})$
be the signature of irreducible representation and
\beq\label{26}
k_{i}= m_{i} +N-i
\eeq
Then eigenvalues of Casimirs are:
\beq\label{27}
c_{n}(\alpha ) = \sum_{i=1}^{N} k_{i}^{n}\gamma_{i}
\eeq
where
\beq\label{28}
\gamma_{i} = \prod_{i\neq j} \left(1-\frac{1}{k_{i}-k_{j}}\right )
\eeq
This is the rather ugly expression. From this it is hard to see (but,
nevertheless, it is true) that $c_{n}(\alpha )$ are {\it symmetric
polynomials} of$\{k_{i}\}$ since generating function for eigenvalues
\beq\label{29}
c(\alpha ,z) \equiv \sum_{n=0}^{\infty}c_{n}(\alpha )z^{n} =
\sum_{i=1}^{N}\frac{\gamma_{i}}{1-k_{i}z}
\eeq
can be represented in the form
\beq\label{30}
c(\alpha ,z) = \frac{1}{z}[1-\pi(\alpha ,z)]
\eeq
where
\beq\label{31}
\pi (\alpha ,z) = \prod_{i=1}^{N}\left(1 - \frac{z}{1-k_{i}z}\right)
\eeq
{}From the last expression the symmetricity and polynomiality of
$c_{n}(\alpha )$ w.r.t. $\{k_{i}\}$ is obvious.

Now let us use this way of \cite{Zhelobenko} in order to derive very "simple"
expression for eigenvalues. From (\ref{31})
\beq\label{32}
\pi (\alpha ,z) = \prod_{i=1}^{N}\frac{1-(k_{i}+1)z}{1-k_{i}z}=
\exp \{\sum_{m=0}^{\infty}\frac{z^{m}}{m}\sum_{i=1}^{N}
[k_{i}^{m}-(k_{i}+i)^{m}]\} = \sum_{n=1}^{\infty}z^{n}P_{n}(\tilde T)
\eeq
where
\beq\label{33}
\tilde T_{n} = \frac{1}{n}\sum_{i}[k_{i}^{n}-(k_{i}+1)^{n}
\eeq
and thus from (\ref{30})
\beq\label{34}
c_{n}(\alpha )= -P_{n+1}(\tilde T)
\eeq
Examples:
\beq\label{35}
c_{0}(\alpha ) = N
\eeq
\beq\label{36}
c_{1}(\alpha ) = \sum_{i}k_{i} - \frac{1}{2}N(N-1) = \sum_{i}m_{i}
\eeq
\beq\label{37}
c_{2}(\alpha ) = \sum_{i}k_{i}^{2} - (N-1)\sum_{i}k_{i} -
\frac{1}{6}N(N-1)(N-2) = \sum_{i}m_{i}^{2} + \sum_{i<j}(m_{i}-m_{j})
\eeq
etc. These expressions (as the Shur polynomials) have the very nice
form in the exponential factor in the bilinear
combination of characters. But, unfortunately, the standard Casimirs
$c_{n}$ with $n>2$ are not factorizable in the simple product form. This is
an obstacle to turn to the determinant form
and, as the consequence, to conclude that our object is a $\tau$-function.
Thus, we propose again to use new Casimirs
like $\sum k_{i}^{n}$ in order to reproduce (\ref{23}) (if we restrict
ourself with $"c_{0}+c_{1}+c_{2}"$ - there is no difference between
these two different definitions, of course). It is in complete analogy with
using only terms of the type ${\rm tr} X^n$, but not ${\rm tr} X^n {\rm tr}
X^m$ in GKM (\ref{gkm}).


\begin{thebibliography}{10}

\bibitem{Douglas} M.Douglas {\sl
  Phys.Lett.}, {\bf 238B} (1990) 176

\bibitem{GMMMO}  A.Gerasimov et al. {\sl Nucl.Phys.}, {\bf B357} (1991) 565

\bibitem{KMMMZ} S.Kharchev et al. {\sl
  Phys.Lett.}, {\bf 275B} (1992) 311;
  {\sl Nucl.Phys.}, {\bf B380} (1992) 181

\bibitem{Mar93} A.Marshakov, {\sl Int.J.Mod.Phys.,} {\bf A8} (1993) 3831

\bibitem{Mor93} A.Morozov, {\it Integrability and matrix models},
Preprint ITEP-M2/93

\bibitem{KazMig} V.Kazakov, A.Migdal, {\sl Nucl.Phys.}, {\bf B397} (1993) 214

\bibitem{IZ} C.Itzykson and J.-B.Zuber, {\sl J.Math.Phys.}, {\bf 21} (1980) 411

\bibitem{KM} S.Kharchev, A.Marshakov, {\it On $p-q$ duality and explicit
solutions in
$c \leq 1$ $2d$ gravity models},
Preprint~NORDITA-93/20, FIAN/TD/04-93

\bibitem{Mir} A.Mironov, {\it On GKM description of multi-criticality in
2d gravity}, Preprint FIAN/TD-17/92

\bibitem{LGGKM} S.Kharchev et al., {\sl Mod.Phys.Lett.,}
{\bf A8} (1993) 1047-1061

\bibitem{Losev} A.Losev, {\it Descendants constructed from matter field
in topological Landau-Ginzburg theories
coupled to topological gravity,} Preprint ITEP, hep-th 9211090;

A.Losev, I.Polyubin, {\it On Connection between Topological Landau-Ginzburg
Gravity
and Integrable Systems}, Preprint UUITP-2/93, hep-th 9305079

\bibitem{Mig}   A.  Migdal, {\sl Zh.Eksp.Teor.Fiz. } {\bf  69
}, 810 (1975) (Sov.Phys.JETP.  {\bf 42} 413)

\bibitem{Rus} B. Rusakov, {\sl
Mod.Phys.Lett.}, {\bf A5}, 693 (1990)

\bibitem{Gross} D.Gross, W.Taylor, {\it Twists and Wilson loops in the
string theory of two dimensional QCD}, Preprint CERN-TH.6827/93, PUPT-1382,
LBL-33767, UCB-PTH-93/09

\bibitem{NeGo} A.Gorsky, N.Nekrasov, {\it Hamiltonian systems of Calogero type
and two dimensional Yang-Mills theory}, Preprint UUITP-6/93, ITEP-20/93

\bibitem{Poly} J.Minahan, A.Polychronakos, {\it Equivalence of Two Dimensional
QCD and the $c=1$ Matrix Model}, Preprint CERN-TH 6843/93, UVA-HET-93-02

\bibitem{Doug} M.Douglas, {\it Conformal Field Theory Techniques for Large N
Group Theory}, Preprint RU-93-13, NSF-ITP-93-39

M.Douglas, V.Kazakov, {\it Large $N$ Phase Transition in Continuum QCD$_2$},
Preprint LPTENS-93/20, RU-93-17

\bibitem{CAMP} M.Caselle, A.D'Adda, L.Magnea and S.Panzeri, {\it Two
dimensional QCD is a one dimensional Kazakov-Migdal model}, Preprint DFTT-15/93

\bibitem{DH} J.Distermaat, G.Heckmann, {\sl Inv.Math.}, {\bf 69} (1982) 259

M.Blau, E.Keski-Vakkuri, A.Niemi, {\sl Phys.Lett.}, {\bf B246} (1990) 92

E.Keski-Vakkuri, A.Niemi, G.Semenoff, O.Tirkkonen, {\sl Phys.Rev.}, {\bf D44}
(1991) 3899

A.Hietamaki, A.Morozov, A.Niemi, K.Palo, {Phys.Lett.}, {\bf B263} (1991) 417

A.Morozov, A.Niemi, K.Palo, {\sl Phys.Lett.}, {\bf B271} (1991) 365; {\sl
Nucl.Phys.}, {\bf B377} (1992) 295

\bibitem{KMSW} I.Kogan, A.Morozov, G.Semenoff, N.Weiss, {\sl Nucl.Phys.},
{\bf B395} (1993) 547

\bibitem{Migd} A.Migdal, {\sl Mod.Phys.Lett.}, {\bf A8} (1993) 359

\bibitem{AFS} A.Alekseev, L.Faddeev, S.Shatashvili, {\sl J.Geom.Phys.},
{\bf v.3} (1989)

\bibitem{Mor} A.Morozov, {\it Mod. Phys. Lett.} {\bf A7} (1992) 3503

\bibitem{Shat} S.Shatashvili, {\it Comm. Math. Phys.} {\bf 154} (1993) 421

\bibitem{Zhelobenko} D.P.Zhelobenko, {\sl Compact Lie group and their
representations,} Nauka, Moscow, 1977

\bibitem{Wit}  E.  Witten, {\sl Comm.  Math.  Phys.}, {\bf
141}, 153 (1991)

\bibitem{MS} G.Moore, N.Seiberg, {\sl Comm.Math.Phys.,} {\bf 123} (1989) 177

\bibitem{DJKM}  M.Sato, Y.Sato, {\it Soliton equations as dynamical systems
in an infinite-dimensional Grassmannian}, in: {\sl Nonlinear partial
differential equations in applied sciences},
P.Lax, H.Fujita, G.Strang (North-Holland,
Amsterdam, 1982)

E.Date, M.Jimbo, M.Kashiwara, T.Miwa, {\it  Transformation groups for soliton
equations},
RIMS Symp. {\sl ``Non-linear integrable systems -- classical theory and
quantum
theory"} (World scientific, Singapore, 1983)

\bibitem{KMMOZ}  S.Kharchev et al. {\sl Nucl.Phys.}, {\bf B366} (1991) 569

\bibitem{KMMM} S.Kharchev et al. {\sl Nucl.Phys.}, {\bf B397} (1993) 379

\bibitem{Ltw}  D.E.Littlewood, {\sl The theory of group characters and
matrix representations of groups}, Oxford, 1958

\bibitem{SW}  G.Segal, G.Wilson, {\sl Publ.Math. I.H.E.S.}, {\bf 61} (1985) 1

\bibitem{Hir}  See, e.g. Y.Ohta, J.Satsuma, D.Takahashi and T.Tokihiro,
{\sl Prog.Theor.Phys.Suppl.,} {\bf 94} (1988) 210

\bibitem{Kac} V.Kac, {\sl Infinite-dimensional Lie algebras},
Cambridge University Press, Cambridge, 1985

\bibitem{Mar} A.Marshakov, {\it On string field theory for $c \leq 1$},
  Preprint~FIAN/TD/08-92

\bibitem{DiMoPle} R.Dijkgraaf, G.Moore, R.Plesser, {\sl Nucl.Phys.}, {\bf B394}
(1993) 356

\bibitem{HC} Harish-Chandra, {\sl Amer.J.Math.,} {\bf 80} (1958) 241

\bibitem{OP} Olshanetsky, Perelomov, {\sl Phys.Rept.}, {\bf 94} (1983) 313

\bibitem{BK} F.Berezin, F.Karpelevich, {\sl Dokl.Akad.Nauk SSSR}, {\bf 118}
(1958) 9-12

\bibitem{NiemiTi} A.Niemi, O.Tirkkonen, {\it On Exact Evaluation of Path
Integrals}, Preprint  HU-TFT-93-7, UU-ITP 3/93

\bibitem{MMM} A.Marshakov, A.Mironov, A.Morozov,  Phys.Lett. {\bf 274B} (1992)
280

\bibitem{KMU} S.Kharchev, A.Mironov, {\it $\tau$-function of two-dimensional
Toda lattice and quantum deformations of the integrable hierarchies}, Preprint
FIAN/TD-02/93

\bibitem{10}   K.Ueno, K.Takasaki, {\sl Adv.Studies in Pure Math.}, {\bf 4}
(1984) 1

\bibitem{26}   M.Jimbo, T.Miwa, M.Sato, {\sl Holonomic quantum fields},
$I-V:$ Publ.RIMS, Kyoto
Univ.,   {\bf 14} (1978) 223; {\bf 15} (1979) 201, 577, 871, 531

\end{thebibliography}
\end{document}